\documentclass[aip,rsi,reprint]{revtex4-1}
\usepackage{graphicx}
\usepackage{dcolumn}
\usepackage{bm}
\usepackage[utf8]{inputenc}
\usepackage[T1]{fontenc}
\usepackage{mathptmx}
\usepackage{etoolbox}
\usepackage{pict2e}
\usepackage{circuitikz}
\usepackage{amsmath,amssymb}
\usepackage{framed}
\usepackage{overpic}
\usepackage{xcolor}
\usepackage{tabularx}

\makeatletter
\def\@email#1#2{
 \endgroup
 \patchcmd{\titleblock@produce}
  {\frontmatter@RRAPformat}
  {\frontmatter@RRAPformat{\produce@RRAP{*#1\href{mailto:#2}{#2}}}\frontmatter@RRAPformat}
  {}{}
}
\makeatother
\begin{document}

\title[LAPPD Gen II  with system-on-a-ship readout]{Initial Assessment of Second Generation of Large-Area Picosecond Photodetectors with Multi-Channel Systems-on-a-Chip Readout}

\author{V.A. Li}
\homepage{https://neutrinos.llnl.gov}
 \email{vali@llnl.gov}
\affiliation{Lawrence Livermore National Laboratory, Livermore, CA, USA}
\author{O.A. Akindele} \affiliation{Lawrence Livermore National Laboratory, Livermore, CA, USA}
\author{M. Bondin} \affiliation{Lawrence Livermore National Laboratory, Livermore, CA, USA}
\author{S.R. Durham} \affiliation{Lawrence Livermore National Laboratory, Livermore, CA, USA}
\author{J.A. Foot} \affiliation{University of California, Merced, CA, USA}
\author{M.J. Ford} \affiliation{Lawrence Livermore National Laboratory, Livermore, CA, USA}
\author{S.-W. Stradleigh} \affiliation{University of California, Merced, CA, USA}
\date{\today}

\begin{abstract}
We first briefly describe the history and motivation behind Cherenkov and scintillation light detection. We then discuss the instrumentation needed to detect these photons as it applies to both photodetectors and readout electronics. One of the motivations is future large neutrino detectors that could in principle differentiate between Cherenkov and scintillation light if using novel water-based scintillators.
In this paper, we present the first measurements utilizing the second generation of Large Area Picosecond Photodetectors (LAPPDs) in conjunction with commercial system-on-a-chip readouts from Nalu Scientific --- specifically, the High Density System on Chip (HDSoC) and Advanced ASoC Rapid Digitizer, Variable Adaptive Readout Chip (AARDVARC) platforms. These state-of-the-art full-waveform digitizers feature sampling rates on the order of 1 and 10 samples per nanosecond, respectively.
Using a picosecond laser, we measured the timing jitter between a pair of LAPPD channels, demonstrating the potential of this setup for precise timing applications.
\end{abstract}

\maketitle

\section{Towards Detecting Scintillation and Cherenkov Light with Picosecond Timing}

This report covers the paradigm of particle detection using light produced by electromagnetic particles in a liquid media, detected by photosensors, and recorded by readout electronics. This paradigm has been proposed to be used in a new neutrino detector (Theia) for the fourth DUNE cavern\cite{Askins}. 

Scintillators and Cherenkov-based detectors have played pivotal roles in some of the most groundbreaking discoveries in particle physics. Notably, Rutherford's early experiments utilized zinc sulfide to detect radiation, marking a significant advancement in the field. Cherenkov's observations of bluish light emitted from radiation-exposed liquids further expanded our understanding of these phenomena. In his Nobel lecture, he concluded, {\it "There can be no doubt that the usefulness of this radiation will in the future be rapidly extended."}

The evolution of detection methods has progressed from early tabletop experiments, which relied on human eyes to detect photons, to the development of large-scale neutrino telescopes such as Super-Kamiokande~\cite{Fukuda2003} and IceCube~\cite{IceCube_Science2023}. These advancements have opened the new field of multi-messenger astronomy. 
The first detection of neutrinos was made possible through scintillator technology~\cite{Science1953}, underscoring the importance of these detection methods in high-energy physics.

In modern detectors, the primary observables are waveforms --- voltages as a function of time --- or integrated charge and timing information from photosensors. These photosensors,  positioned within the detector, register light generated by particle interactions. The recorded charge, timestamps, and spatial coordinates of the photosensors are then processed to extract meaningful information, enabling the identification and reconstruction of the underlying physics events.

\subsection{Generation of scintillation and Cherenkov light}
Discoveries of both scintillation and Cherenkov radiation were made before a photomultiplier tube (PMT) was invented. Both discoveries were made by visually observing photons through a microscope~\cite{crookes1903emanations,Cherenkov1934,Vavilov1934}, and later using photography~\cite{PhysRev.52.378}.

When a charged particle is moving in a medium with an index of refraction $n$ higher than the speed of light ($v > c/n$) in that medium, it emits photons.
These photons will be emitted instantaneously with the particle motion at an angle $\theta_C = \arccos (c / (vn))$; for ultra-relativistic particles ($v \simeq c$), it is $\theta_C = \arccos (1/n)$.
For water ($n = 1.33$), it is $\sim 41^\circ$, for organic scintillators ($n \simeq 1.5$) $\sim 49^\circ$.
The number of emitted photons per unit length and unit wavelength interval could be estimated by Frank-Tamm equation:
\begin{equation}
\frac{d^2 N}{dx d\lambda} = 
\frac{2\pi \alpha}{\lambda^2} 
\left(1 - \frac{1}{\beta^2 n^2(\lambda)}\right)
\end{equation}
Although the emission is peaking at lower energies (due to $1/\lambda^2$ dependence above), 
the majority of the photons are in the visible range since photons with shorter wavelengths are absorbed before reaching the photosensor.
As an example, one could estimate for water in the [400, 500]-nm range, after integrating the equation,  it yields about 100 photons per centimeter.

Scintillators, made of materials such as plastics, organic liquids, or crystals, fluoresce light in the UV to visible range through atom excitation and emission when absorbing ionizing radiation. Compared to Cherenkov radiation, which emits photons purely on particles moving faster than light in a medium, scintillators provide better energy resolution due to the generally higher light yield but lose directional information.

A scintillator is typically characterized by its light output and the time it takes for scintillation light to be emitted after the passage of a charged particle. The fastest scintillators are those with very short decay times, allowing them to respond quickly to radiation events. Organic scintillators (e.g., EJ-200) are among the fastest, with decay times of 1--2 nanoseconds, while inorganic scintillators (e.g., NaI, BGO) are usually slower, with decay times on the order of tens to hundreds of nanoseconds. For scintillation light yield, it depends on the type of scintillator, for some liquid and plastic scintillators it is of the order of 10,000 photons per MeV of electron equivalent energy.
The quenching is different for different particle types, and thus is reported in unit of electron equivalent energy. 
The wavelength at which the scintillation light is emitted generally peaks in the (400, 450)-nm range for organic scintillators. 
The light is isotropically emitted as the fluor molecules de-excite.
Table~\ref{tab_scint_Cherenkov} highlights the main differences between scintillation and Cherenkov light.

In general, for ``pure'' organic scintillators, the scintillation light is two orders of magnitude brighter than Cherenkov light.
It was therefore of interest to conceive a new class of materials called water-based scintillators, 
where the ratio between scintillation and Cherenkov photons is optimized.
The primary purpose is lowering the cost of the detection medium (for large neutrino detectors).

\begin{table}[]
    \centering
    \begin{tabular}{|l|c|c|}\hline
    &  Cherenkov & Scintillation\\\hline
        Spectrum & $\sim 1 / \lambda^2$  & Poisson-like \\
        Timing & Prompt & Delayed \\
        Direction & $\cos \theta \sim 1 / n$ & Isotropic \\\hline
    \end{tabular}
    \caption{General differences between scintillation and Cherenkov light --- spectrum, timing, direction.}
    \label{tab_scint_Cherenkov}
\end{table}

\begin{figure}
    \centering
    \includegraphics[width=1\linewidth]{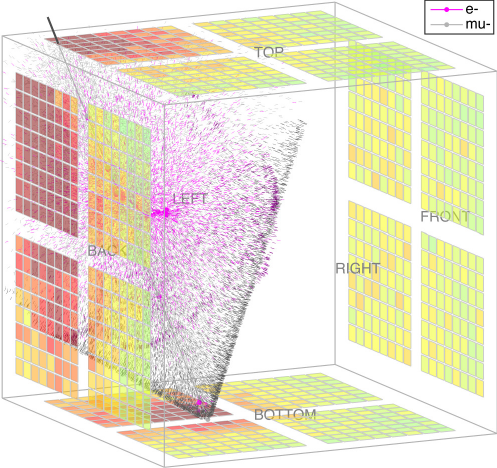}
    \caption{
    Cherenkov photons and the need for fast timing — an example demonstrating the uniform and directional nature of Cherenkov photons (shown in gray). Magenta photons represent those emitted by knock-on electrons. This simulation also highlights the need for fast timing electronics and fine-grained photosensors in a pixelated detector. The simulation depicts a cosmogenic muon with a few GeV of energy traversing a plastic scintillator cube (13-cm side), producing a knock-on electron that excites the scintillator molecules. The snapshot is taken approximately 1~ns after the muon entered the cube through the top face and is now exiting through the bottom face. 
The relativistic muon, traveling faster than the speed of light in the medium, produces Cherenkov photons along its path at an angle of approximately 42 degrees to the track. The detector in this example, the miniTimeCube, was tiled with Planacon MCP-PMTs; however, one could imagine using LAPDs instead. At this point in the simulation, scintillation photons have not yet been emitted due to the 1–2 nanosecond delay after the particle's passage. Figure adapted from \cite{mTC:2016yys} (V.A. Li et al., Review of Scientific Instruments, Vol. 87, 021301, 2016; licensed under a Creative Commons Attribution (CC BY) license.)}
    \label{fig_mTC_scint_Cherenkov}
\end{figure}

We can further consider a 1-MeV $\beta$-ray (either positron or electron), it will emit Cherenkov light until reaching $\sim$260-keV kinetic energy or for about 2-3~mm in water (assuming stopping power of 2 MeV cm$^{-2}$ g), it will emit about 20-30 photons. 
At the same time, in water-based liquid scintillator (WbLS) or pure liquid scintillator (LS) it will emit somewhere between 100 and 10,000 scintillation photons, depending on the water/scintillator ratio \cite{Yeh:2011zz}.
The Cherenkov light production based on a simulation is illustrated in Fig.~\ref{fig_mTC_scint_Cherenkov}.

Unlike Cherenkov photons, scintillation light is not emitted in a preferred direction (as it is emitted by the excited triplet states, not by the particle itself).
As a result, there is also a delay of emission for scintillation light compared to Cherenkov radiation. For fast scintillators, it is on the order of nanoseconds, but it can be longer for slow scintillators. The slight timing difference allows for the separation of Cherenkov and scintillation-emitted photons in photodetectors, although limitations in their timing resolution and transit-time spread of single photoelectrons restricted experimental measurements until recently, with discrimination by LAPPDs first proposed in 2014\cite{Aberle_2014}.
In pure liquid scintillator, the majority of photons are scintillation.
In pure water, there are no scintillation photons, only Cherenkov photons.

Naturally, WbLS allows ``separation'' between scintillation and Cherenkov photons.
In WbLS, it is on the order a few nanoseconds~\cite{Yeh:2011zz}.
WbLS can also be made pulse-shape sensitive, allowing differentiation between electronic and nuclear recoils based on the waveform shape~\cite{Ford:2022wla}. Recent experiments have also demonstrated this ability to perform scintillation and Cherenkov photon discrimination in WbLS using a PMT array and/or LAPPDs by the CHESS experiment, ANNIE group, and Kaptanoglu, et al\cite{Caravaca_2020,anniecollab2024,Kaptanoglu_2022}.

Another method of discriminating scintillation and Cherenkov photons uses wavelength separation in certain WbLS where the scintillator photon wavelengths do not significantly contaminate Cherenkov photon wavelengths\cite{Kaptanoglu_2019}. 
Experimentally, this can be performed through chromatic separation using dichroic filters shaped into a Winston-style light concentrator with photodetectors (termed 'dichroicons'), and was recently demonstrated. This allows the wavelengths of Cherenkov and narrow-band scintillation photons to be individually distinguished and sorted to their corresponding wavelength-sensitive photodetector instead of relying on timing-based methods, and has the benefit of improved directional information from dispersion measurements\cite{Kaptanoglu_2020}. 

\subsection{Detection of light --- conversion to electrons}

In the 1930s, the development of the first PMT and the cathode ray tube (CRT) significantly advanced technologies such as oscilloscopes and televisions \cite{Kubetsky,Zworykin}. The invention of PMTs facilitated numerous scientific discoveries by enabling the detection of scintillation and Cherenkov radiation. Additionally, the integration of arrays of PMTs has enhanced the observation of detection media, paving the way for advancements in particle physics, astrophysics, and medical imaging.

At the core of photon detection lies the photoelectric effect, which involves the conversion of photons into electrons, commonly referred to as photoelectrons. This led to the introduction of materials known as photocathodes in the 1920s, which are based upon metals that exhibit the photoelectric effect and were quickly used to build the first PMTs. Photocathodes are characterized by their quantum efficiency, work function, lifetime, optical properties, and timing response. They were initially constructed out of simple metal plates, but a focus on improving their characteristics led to further developments into metalloid crystal materials such as semiconductors.

PMTs operate through the basis of amplification, where a weak signal produced by a single incoming photon is amplified to levels detectable by typical electronics. This amplification process is achieved through electrodes in a vacuum known as dynodes. An applied electric field accelerates incoming electrons towards each dynode, causing secondary emissions of photoelectrons that continually multiply them to cause a process known as an electron avalanche.
A typical PMT achieves a gain of \(10^6\) to \(10^7\), meaning that a single photoelectron (PE) can be amplified to produce millions of electrons at the output. This amplification corresponds to a charge of approximately \(1.6 \times 10^{-12}\) C. This charge, typically spread over a few nanoseconds, corresponds to a detectable current of about 1~mA.

The concept of continuous dynodes is foundational to the development of micro-channel plates (MCPs). The idea is that gain comes from electrons colliding with the walls of the microchannels causing secondary emissions. So each microchannel act like a tiny continuous dynode which are typically $\sim 10\; \mu$m in size. The first micro-channel plate designs emerged in the 1960s, introducing the idea of continuous electron multiplication through the use of plates composed of numerous micro-channels~\cite{Oschepkov1960,Goodrich1962,Timothy2013}.
Similar to CRTs, MCPs can be utilized for the detection of both photons and charged particles when operated in a "reverse" configuration. This versatility has made MCPs an important tool in various detection applications.
The history of micro-channel plates is well-documented, highlighting their evolution and significance in the field of particle detection~\cite{Timothy2013}.

Parallel advancements in electronics for instruments that specialize in fast waveform sampling and affordable multi-channel electronics have further enabled multi-anode MCP-PMT (MA MCP-PMT) development due to the large increase of information that needs to be processed in high-energy physics. Neutrino and accelerator experiments have revealed demand for large-area coverage detectors using MA MCP-PMTs to obtain a higher probability of event detection and correlated imaging through high spatial resolution. High temporal resolution is also desired from MCP-PMTs for event characterization that distinguishes between fast-timing events separated by small orders of magnitude at sub-nanosecond time scales. As noted in the previous section, Cherenkov and scintillation light differ in their timing, intensity, and spatial emission; a neutrino experiment utilizing a photodetector that has these characteristics is ideal for separation and reconstruction of such photons along with financial benefits in replacing immense arrays of traditional PMTs \cite{Adams2016}.

\begin{figure}
    \centering
    \includegraphics[width=1\linewidth]{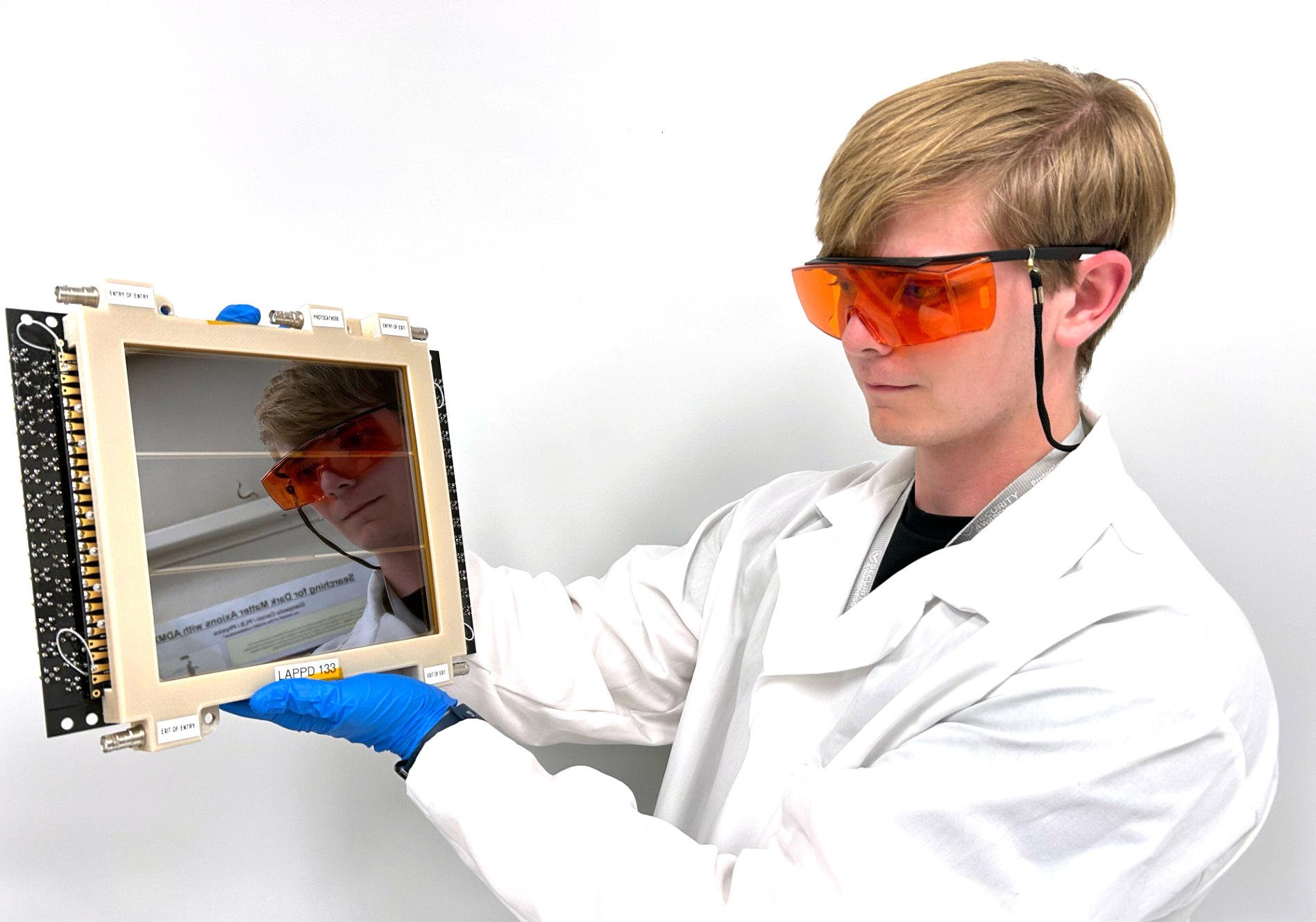}
    \caption{Large-area picosecond photo detector. A photograph of an LAPPD Gen 2, with a person holding it shown to scale. An active square area show is 20~cm $\times$ 20~cm. The two support ``ribs'' are noticeable, as well as five SHV connectors; the SMA signal connectors are at the back.}
    \label{fig_LAPPD_James}
\end{figure}

MA MCP-PMTs developed by companies such as Burle and Hamamatsu in the 1990s have paved the way for large-area MA MCP-PMTs. In 2006, Inami et al. demonstrated high temporal resolution of <10~ps from Cherenkov light, establishing MCP-PMTs as a leading candidate for future ultrafast photodetectors \cite{INAMI2006}. The inherently small pore-geometry also naturally provides advantages for high two-dimensional spatial resolution \cite{Wiza1979}. The Large Area Picosecond Photodetector (LAPPD) gen 1 is one such MA MCP-PMT \cite{LAPPD_gen1}, which is now accessible for various applications. Further evolution has produced LAPPD gen 2, which is a type of detector that, strictly speaking, is not a multi-anode MCP-PMT while still containing a pixelated readout (shown in Fig.~\ref{fig_LAPPD_James}, is the LAPPD gen 2 used in this study). The first-generation LAPPD was first characterized using pulsed sub-picosecond lasers to determine a time resolution of <70~ps, a spatial resolution of $\sim$500~$\mu$m, and MCP gain above $10^7$ for photoelectrons \cite{Adams2015}. LAPPDs are now being used in physics measurements and recently demonstrated their capability to observe neutrinos in the ANNIE experiment \cite{ANNIELAPPD}.

\subsection{Recording Electronic Signals: Waveform Digitizers}

The field of multi-channel waveform digitizers is currently grappling with the challenge of efficiently handling high-channel density photosensors. The desire is to have precise timing to separate Cherenkov and scintillation light across multiple detectors. For a future neutrino detector, it will become needed to be capable to read all pixels from detectors at any given point, such as the 64 channels per LAPPD. Although oscillographs, the precursors to oscilloscopes, existed prior to the advent of the CRT, the introduction of CRT technology greatly enhanced their widespread adoption, with digital oscilloscopes transforming the ability to capture and analyze waveforms at high densities. 

Progress in complementary metal-oxide-semiconductor (CMOS) technology have led to the development of field transistors with remarkably small sizes, now reaching down to 20~nm. The timeline of application-specific integrated circuits (ASICs) and system-on-chip (SoC) technologies reflects the rapid progress in instrumentation, where engineering innovations are driven to the limits of device miniaturization by the principles of quantum mechanics, leading to numerous groundbreaking discoveries.

A fundamental principle in both analog and digital oscilloscopes is the linear relationship between voltage and time, rooted in the Time to Amplitude Converter, which can be expressed as:

\begin{equation}
    \Delta V = \frac{\Delta q}{C} = \frac{\Delta q}{\Delta t} \frac{\Delta t}{C} = \frac{I}{C} \Delta t
\end{equation}

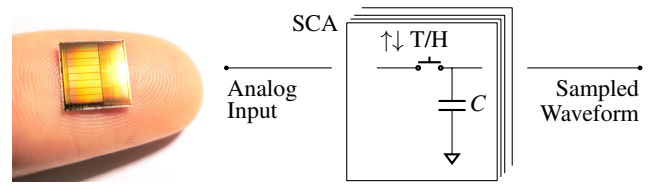
\begin{figure}[ht]
\begin{tabular}{cc}
\includegraphics[width=.3\linewidth]{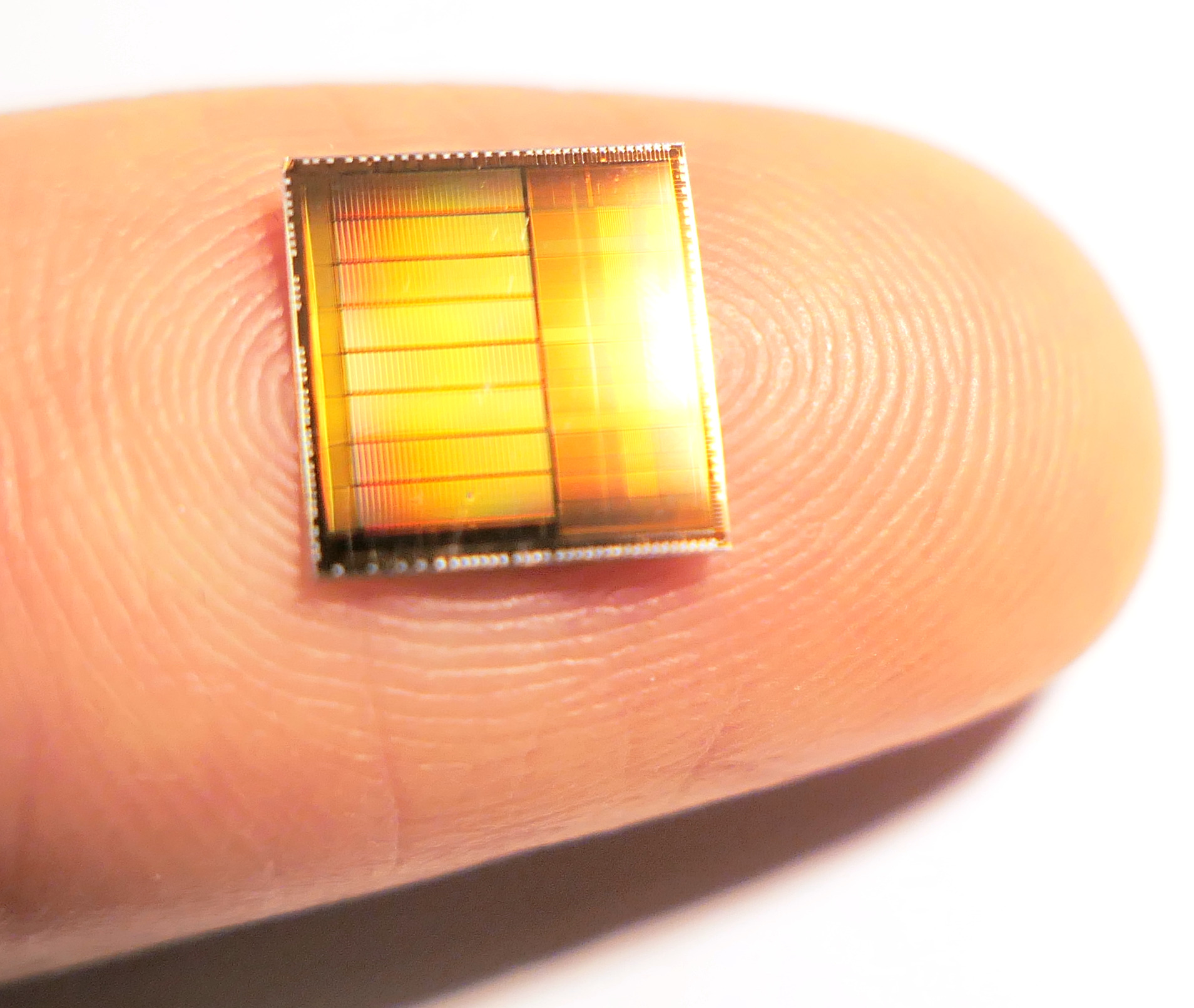}
&
\begin{circuitikz}
\draw
  (-0.1, 0.7) node[right]{Analog}
 (-0.1, 0.4) node[right]{Input}
  (0,1.0)  to [short, *-] (1.4,1)
  
  (4,1) to [short, -*] (5.5,1) 
  (5.6,0.7) node[left]{Sampled} 
    (5.6,0.4) node[left]{Waveform} 

  (2,1) to [push button] (3.4,1) 
  (2.5,1.1) node[above]{$\uparrow \downarrow$ T/H}
  (3,0) node[sground]{} to [C, l_=$C$] (3,1)
  (1.6,-0.5) to [short, -] (1.6,1.6)
  (3.6,1.6) to [short, -] (1.6,1.6)
  (3.6,-0.5) to [short, -] (3.6,1.6)
  (3.6,-0.5) to [short, -] (1.6,-0.5)

  (3.8,1.8) to [short, -] (1.8,1.8)

  (3.7,1.7) to [short, -] (1.7,1.7)
  (3.65,1.65) to [short, -] (1.65,1.65)

  (3.65,-0.45) to [short, -] (3.65,1.65)
  (3.7,-0.4) to [short, -] (3.7,1.7)

  (3.8,-0.3) to [short, -] (3.8,1.8)

  (1.6,1.6) node[left]{SCA}
; 
\end{circuitikz}
\\
\end{tabular}
\caption{Digitization concept --- track and hold  --- the underlying idea behind the Switched Capacitor Array (SCA). 
Thousands or tens of thousands of capacitors per channel, and multiple channels within an ASIC/SOC.
A photograph of a typical die (without the plastic package) --- in this example --- an 8-channel Ice Ring Sampler (IRS), a predecessor to some of the Nalu chips, in addition to capacitors, the chip features several millions of transistors and resistors.} 
\label{fig_SCAs_IRS_die}
\end{figure}

Historically, the desire to record waveforms dates back to the advent of alternating current, with early methods using switches and condensers. Significant advancements in sub-nanosecond and sub-10 picosecond timing have been made, with notable contributions from the late Professor Gary Varner at the University of Hawaii and the Instrumentation and Development Laboratory, now VarnerLab. This includes a family of ASICs, such as Labrador, BLAB, IRS (shown in Fig.~\ref{fig_SCAs_IRS_die}), and TARGET, embodying the concept of a ``digital oscilloscope on a chip''\cite{Ruckman:2008cj}. An underlying idea is the track and hold operation\cite{Horowitz, Baker2019}. Each ASIC typically includes millions of transistors and resistors, and it usually takes several years to design and test it. 

The terminology surrounding ASICs and SoCs can be ambiguous; ASICs are custom-designed for specific challenges and lack the general computational capabilities of microprocessors, DSPs, FPGAs, or microcontrollers. Most SoCs are a type of ASIC, inherently heterogeneous with various partitions performing distinct functions and a central digital component managing the system. Moreover, they are based on switched-capacitor-array (SCA) architectures. The idea is for capacitors and switches to perform precise charge transfer without the need of resistors. This allows for high accuracy, lower power consumption, and compactness. Some issues with SCA architecture design come from the overall design complexity, integration, and deadtime. The DRS4 and PSEC4 architecture are some such examples\cite{Stefan2010, Oberla2014}.

In our initial readout tests for the LAPPD, we utilized AARDVARC, Advanced ASoC Rapid Digitizer, Variable Adaptive Readout Chip, and the HDSoC, High-Density System-on-chip, from Nalu Scientific for readout. AARDVARC is a 4 channel digitizer developed in 2019 that runs at 10 Gigasamples per second. While HDSoC is a 32 channel digitizer developed in 2022 that runs at 1 Gigasamples per second. Both cards were designed to be able to readout data with good timing accuracy and high data transfer rates, with HDSoC designed for High Density Detectors. For our initial readouts, we found that the faster sample rate of the AARDVARC was more desirable since we tested on a couple of pixels at a time.

Both AARDVARC and HDSoC resemble Gary Varner's ``oscilloscope on a chip'' but include additional digital control and processing features to streamline data acquisition management. Notably, the AARDVARC chip includes a small CPU for self-calibration tasks and complex autonomous acquisition processes, although this feature is not yet fully publicized due to ongoing testing. In principle, an FPGA can be integrated into AARDVARC, as it has some FPGA resources available~\cite{private_communication_Luca_Nalu}.

\ctikzset{bipoles/length=.6cm}

\section{Experimental setup: LAPPD gen 2}

\begin{figure}
    \centering
    \includegraphics[width=0.49\linewidth]{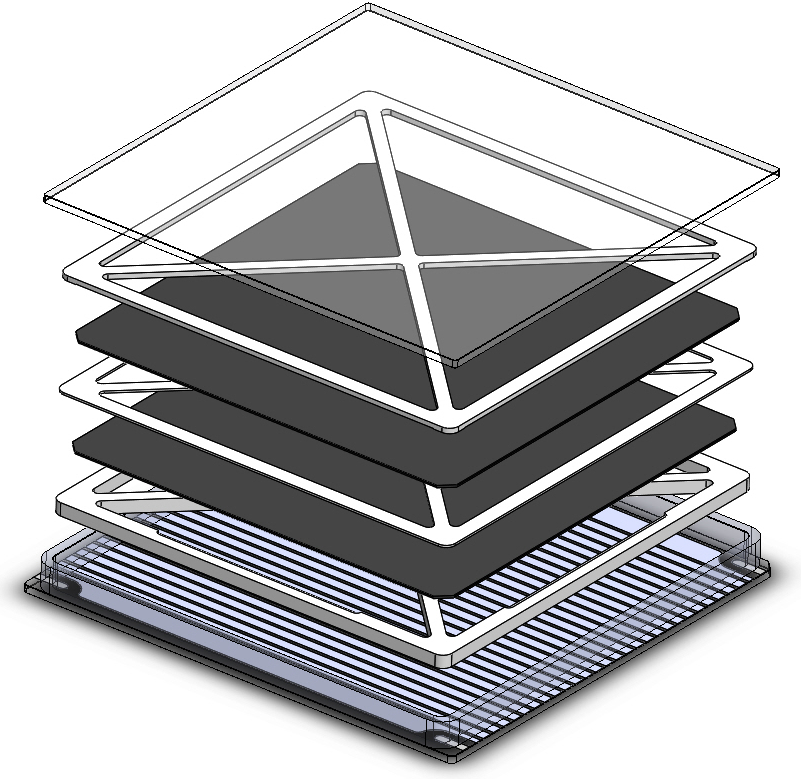}
    \includegraphics[width=0.49\linewidth]{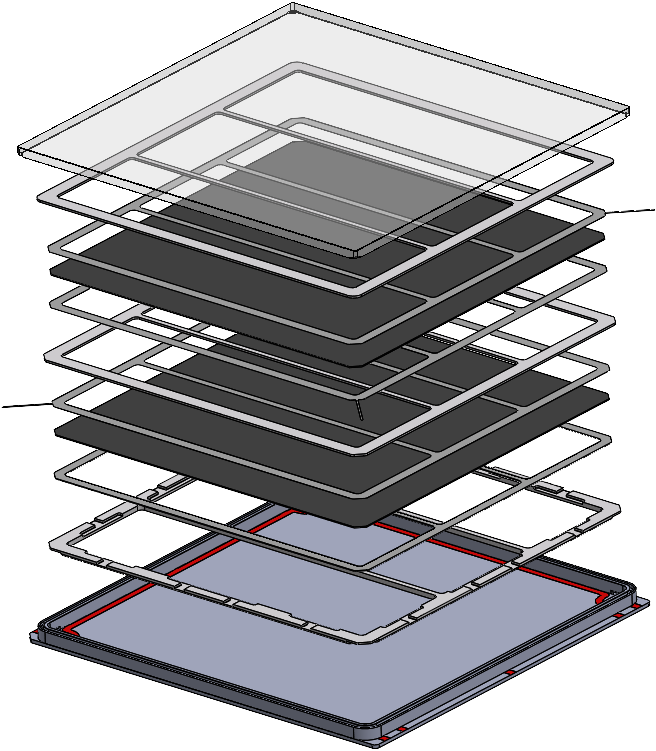}

       \vspace{2mm}
\begin{tabularx}{\linewidth}{X X}
    \centering Gen 1 & \centering Gen 2 
\end{tabularx}
       \vspace{-5mm}
    \caption{
    Exploded view of first- and second-generation LAPPD modules. In the first generation, spacers are arranged in a cross pattern and the anode consists of parallel strips. In the second generation, spacers are two parallel bars and the anode is a single resistive plane. In both diagrams, the top plane represents the glass window with deposited photocathode; three spacers support the glass and two MCPs (MCPs are shown in dark gray); the anode plane is shown in light gray-blue. In the second generation, shims surrounding each of the two MCPs are also depicted. Figures reproduced with permission from Incom.}
    
    \label{fig_LAPPD_gen_1and2}
\end{figure}

\begin{figure}
    \centering
    \begin{overpic}[width=1\linewidth]{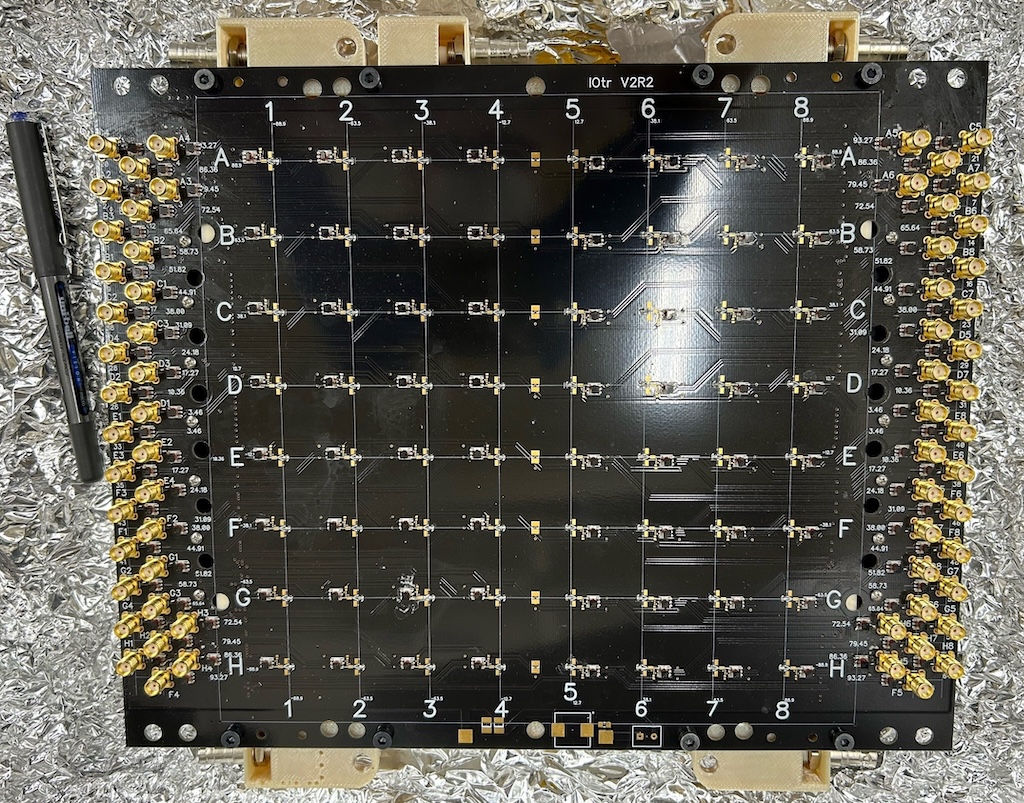} 
        \thicklines
        \put(55,74){\fcolorbox{red}{white!30}{\parbox{15pt}{$V^{PC}$}}}
        \put(55,74){\color{red}\vector(-1,0){4}}
        \put(90,74){\fcolorbox{red}{white!30}{\parbox{15pt}{$V_T^T$}}}
        \put(90,74){\color{red}\vector(-1,0){4}}
        \put(4,74){\fcolorbox{red}{white!30}{\parbox{15pt}{$V_T^B$}}}
        \put(13,74){\color{red}\vector(1,0){4}}
        \put(6,5){\fcolorbox{red}{white!30}{\parbox{15pt}{$V_B^B$}}}
        \put(15,5){\color{red}\vector(1,0){4}}
        \put(90,4){\fcolorbox{red}{white!30}{\parbox{15pt}{$V_B^T$}}}
        \put(90,4){\color{red}\vector(-1,0){4}}
        \put(40,4){\color{red}\vector(-2,1){22}}
        \put(68,4){\color{red}\vector(2,1){19}}
        \put(40,3){\fcolorbox{red}{white!30}{\parbox{65pt}{\centering \textcolor{black}{ 64 signal outputs}}}}
        \thinlines
    \end{overpic}
    \caption{A photograph of an LAPPD \#133 Gen 2, backplane. Five SHV connectors are: photocathode, entry and exit of the entry MCP, and entry and exit of the exit MCP. 64 SMA connectors are for the anode. 64 transformers are also present at the backplane next to each anode.}
    \label{fig_LAPPD_back}
\end{figure}
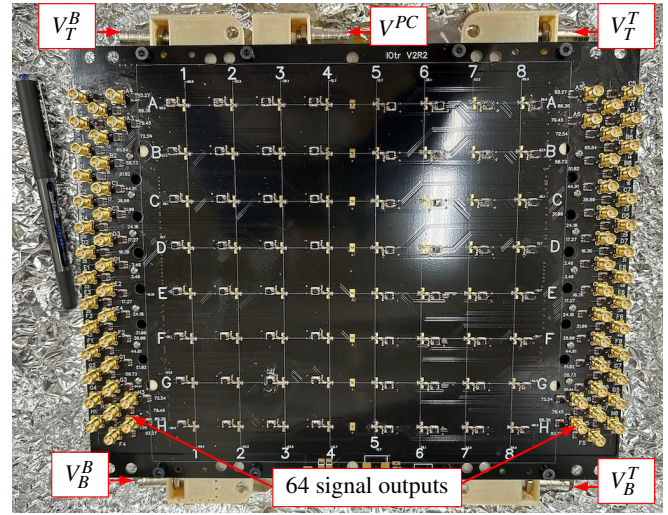

The LAPPD is a multi-anode device with two micro-channel plates inside a glass enclosure under vacuum.
The LAPPD is constructed with a 5.0-mm-thick fused-silica glass window, which provides good light transmission. It uses a bialkali photocathode material (Na$_2$KSb) that operates effectively within a spectral response range of 160 to 650~nm photons for their conversion to photoelectrons. The detector has maximum sensitivity at wavelengths of 365~nm or less, making it suitable for ultraviolet detection. The active area measures 195~mm $\times$ 195~mm, resulting in a minimum effective area of 373~cm$^2$. Additionally, the design includes edge frame rib-spacers, allowing for an active fraction of 97\%, which enhances light collection efficiency, compared to the cross-shaped spacers used in the first generation of devices, shown in Fig.~\ref{fig_LAPPD_gen_1and2}. Two 203~mm $\times$ 203~mm MCPs with a pore size of 20~$\mu m$ cause secondary emissions for electron amplification after the photocathode, with the outgoing electron cloud pulled by the applied electric field to the resistive anode. The anode records the deposited charge signal, which results in an induced current that capacitively couples to pickup electrodes, being an array of conductors, and a signal ground to create an external readout board \cite{Angelico_2017}. In the default external readout board for the LAPPD, a 8 $\times$ 8 pixel pad array acts as the pickup electrode, which feeds signal to 64 SMA outputs, Fig.~\ref{fig_LAPPD_back}.

Photodetector physical dimensions (L $\times$ W $\times$ Thickness, in millimeters) are  230 $\times$ 220 $\times$ 22.
The LAPPD is enclosed in Ultem housing, on which five HV connectors are mounted. 
Overall footprint, with mounting case and printed-circuit interface board  are 300~mm $\times$ 274~mm $\times$~26.8 mm.

The most prominent feature of the second-generation LAPPD\cite{Shin:2022ybc} is its capacitive readout: an internal resistive film anode couples to a customizable external readout board, enabling application-specific pad geometries and segmentation for signal pickup. By contrast, the first generation uses a stripline anode that is sealed inside the vacuum package and read out directly. In addition, Gen 2 replaces the supporting crosses with smaller-area supporting ribs, which reduces detector dead space and increases the active photon-collection area, as shown in Fig.\ref{fig_LAPPD_gen_1and2}; this spacer design has also been adopted in newer Gen 1 devices.

\begin{figure}
    \centering
    \includegraphics[width=1\linewidth]{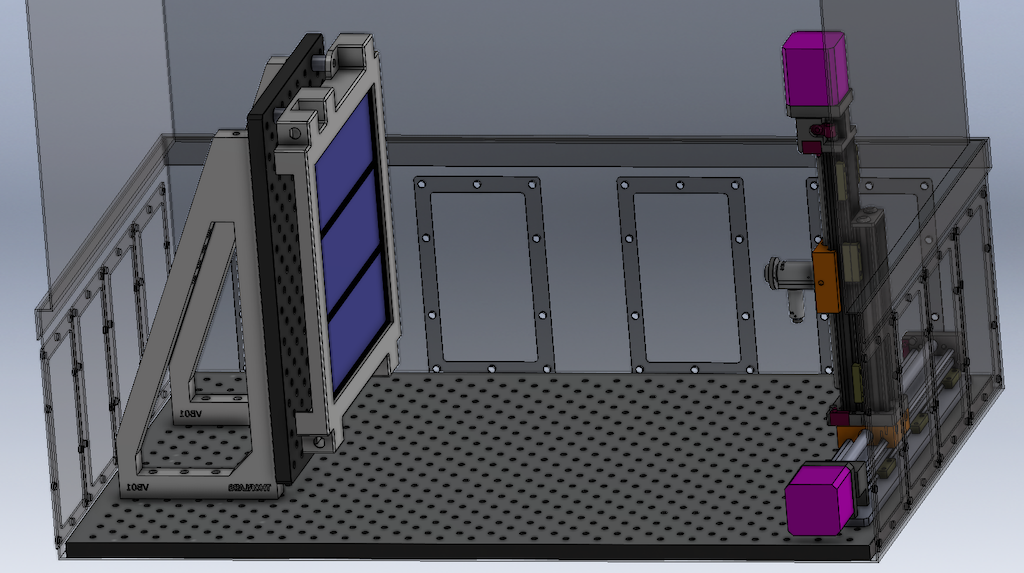}
    \begin{overpic}[width=1\linewidth]{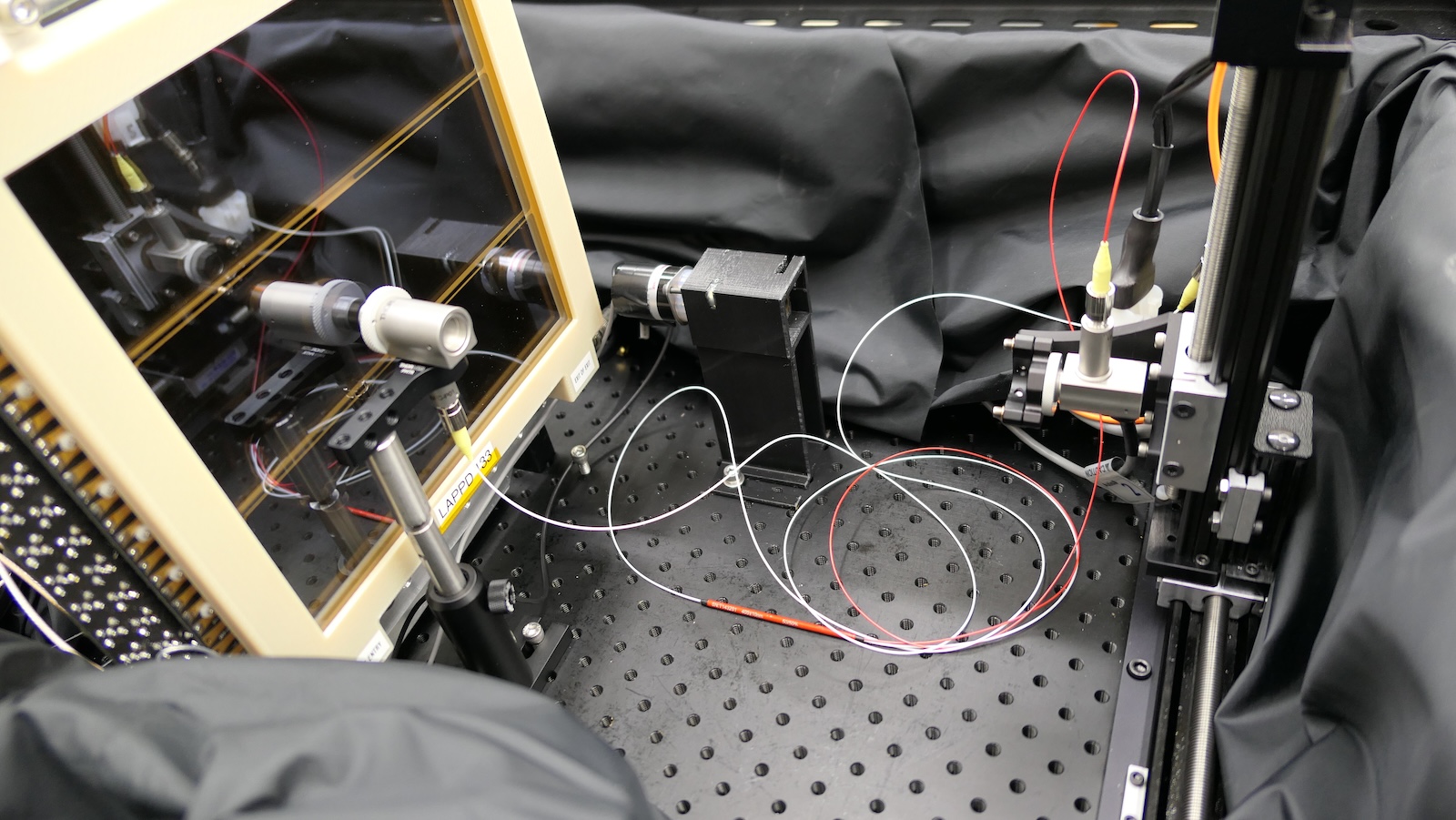} 
        \put(2,52){\fcolorbox{red}{white!30}{\parbox{40pt}{\centering \textcolor{black}LAPPD}}}
        \put(50,40){\fcolorbox{red}{white!30}{\parbox{20pt}{\centering \textcolor{black}PMT}}}
        \thicklines

        \put(27,10){\color{white}\vector(0,1){22}}
        \put(50,7){\color{white}\vector(0,1){8}}
        \put(75,10){\color{white}\vector(0,1){19}}
        \put(42,5){\fcolorbox{red}{white!30}{\parbox{40pt}{\centering \textcolor{black}Splitter}}}

        \put(65,5){\fcolorbox{red}{white!30}{\parbox{50pt}{\centering \textcolor{black}Collimator on $xy$ stage}}}
        \put(15,5){\fcolorbox{red}{white!30}{\parbox{50pt}{\centering \textcolor{black}Fixed Collimator}}}

        \thinlines
    \end{overpic}
    \caption{A CAD rendering and a photograph of an LAPPD Generation 2 mounted on an optical breadboard inside a light-tight enclosure. The fiber-based 50:50 beam splitter and two collimators allow to shine the light on two different pixels.  A collimator is mounted on a  XY-linear stage, while the other is fixed. A reference 1-inch PMT is also mounted nearby LAPPD without obscuring the view. }
    \label{fig_CAD_LAPPD_new}
\end{figure}
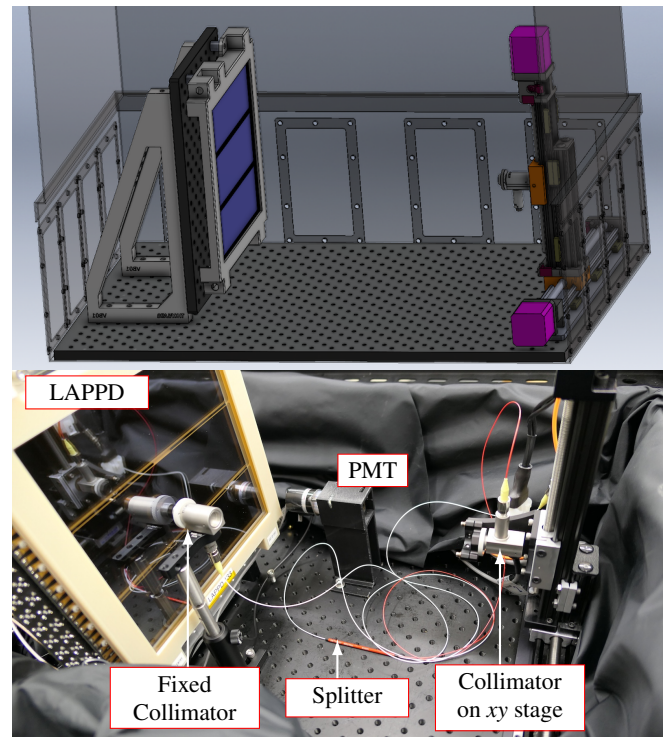

A special light-tight enclosure was constructed. In addition to the LAPPD, optical fibers, a reference PMT, a beam splitter, two optical collimators --- one fixed and one movable in $xy$ plane  parallel to the LAPPD face, on a stepper-motor-driven carriage. The setup is shown in Fig.~\ref{fig_CAD_LAPPD_new}.

\begin{figure}
\begin{tabular}{lcr}
    \begin{circuitikz}
\draw
   (0,5) node[above]{$V^T_T$} to [short, *-*] (2,5)
   (1,5 ) to [R, l=$R_1$] (1,4) 
   (0,4) node[above]{$V^T_B$} to [short, *-] (1,4)
   to [R, l=$R_2$] (1,2) 
   (0,2) node[above]{$V^B_T$} to [short, *-] (1,2)
   to [R, l=$R_3$] (1,1) 
   (0,1) node[above]{$V^B_B$} to [short, *-] (1,1)
   (1,0) node[ground]{} to [R, l_=$R_4$] (1,1)
; 
\end{circuitikz} 
 & \qquad \color{white}{-----------} &
 \begin{circuitikz}
\draw
   (0,5) node[above]{$V^T_T$} to [short, *-*] (2,5)

   (0,4) node[above]{$V^T_B$} to [short, *-*] (2,4)
   (1,3) node[ground]{}  to  [R, l_=3 M$\Omega$] (1,4)
   
   (0,2) node[above]{$V^B_T$} to [short, *-*] (2,2) 
   (0,1) node[above]{$V^B_B$} to [short, *-*] (2,1)
   (1,0) node[ground]{} to [R, l_=0.5 M$\Omega$] (1,1)
; 
\end{circuitikz}
   \\\vspace{2mm}
\end{tabular}
\begin{tabularx}{\linewidth}{X X}
    \centering Method 1 & \centering Method 2 
\end{tabularx}
    \caption{Two connection diagrams of the two methods for biasing two LAPPD MCPs: left panel --- voltage divider (single HV channel is required); right panel --- individual HV channels with exit of each MCP having a resistor to ground. In this paper, we have chosen the second method (with both resistors rated 1\;W).}
    \label{fig_MCP_connect}
\end{figure}
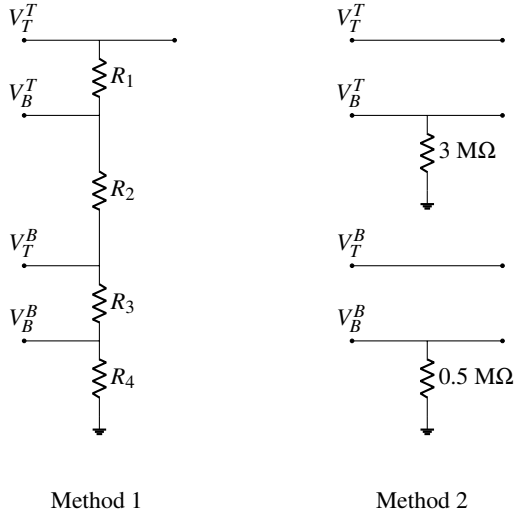

Two different ways of powering up the LAPPD were explored. The two methods are represented in Fig.~\ref{fig_MCP_connect}. The method with the voltage divider requires less channels, but is less versatile as the user needs to ensure each LAPPD has its own voltage divider with resistor values specifically tuned for a particular LAPPD. The method of having five independent high-voltage channels is more reliable as it provides more flexibility of setting voltages. We choose the method of independently providing high voltages to the photocathode ($V^{PC}$), top of top MCP ($V_T^T$), bottom of top MCP ($V_B^T$), top of bottom MCP ($V_T^B$), and bottom of bottom MCP ($V_B^B$). We differ from Incom's convention of Entry and Exit, which have been replaced by Top and Bottom respectively to better distinguish the two points in superscript notation.

\ctikzset{bipoles/length=.6cm}

After SHV cables are connected, the voltages are ramped up in steps to ensure the currents are stable. The dark noise start to appear.
The procedure is demonstrated by Table~\ref{tab_LAPPD_stes_V_and_I}. Gradual turning on of the LAPPD is recommended by the manufacturer. The voltages should be ramping up at about the same rate, all 5 channels at once. At step 3, one could start seeing occasional dark noise. Once the photocathode voltage exceeds the top of the top MCP, the dark rate significantly increases --- there might be multiple steps between steps 4 and 5. The table is an example one could use to troubleshoot an LAPPD if the voltages are unstable. The current values are those measured by the high-voltage supply, and listed in micro-amperes.

\begin{table}[ht]
    \centering
\begin{tabular}{|l||c|c||c|c||c|c||c|c||c|c|}
        \hline
         & \multicolumn{4}{c||}{Bottom MCP} & \multicolumn{4}{c||}{Top MCP} & \multicolumn{2}{c|}{Photocathode} \\ \hline
        Step & $V^B_B$ & $I^B_B$ & $V^B_T$ & $I^B_T$ & $V^T_B$ & $I^T_B$ & $V^T_T$ & $I^T_T$ & $V^{PC}$ & $I^{PC}$ \\ \hline
        1 & 200  & 387  & 600  & 15 & 800  & 256  & 1200 & 12 & 1190 & 0.12 \\
        2 & 200 & 379 & 800 & 23 & 1000 & 316 & 1600 & 18 & 1590 & 0.11 \\
        3 & 200 & 372 & 1000 & 30 & 1200 & 377 & 2000 & 24 & 1990 & 0.08  \\
        4 & 200 & 371 & 1000 & 31 & 1200 & 377 & 2000 & 24 & 2010 & 0.07 \\
        5 & 200 & 371 & 1000 & 31 & 1200 & 377 & 2000 & 24 & 2200 & 0.07 \\ \hline
    \end{tabular}
    \caption{Set voltages (in volts, negative sign omitted) and measured currents (in micro-amperes) across both MCPs and photocathode of the LAPPD. }
    \label{tab_LAPPD_stes_V_and_I}
\end{table}

\begin{figure}
    \centering
    \includegraphics[width=1\linewidth]{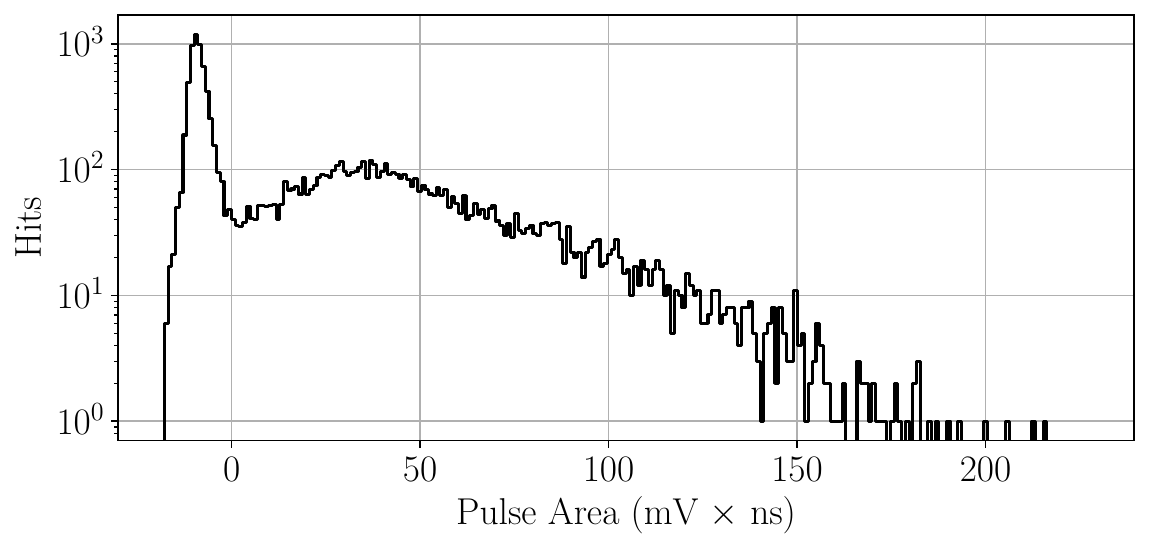}
    \caption{Pulse area histogram for SPE pulses obtained with the oscilloscope. The SPE peak is at about 50~pVs (the baseline is at about $-$10~pVs).}
    \label{fig_SPE_gain_TEKscope_laser_splitter_H1}
\end{figure}

To estimate the gain of a single pixel that an LED or laser (in a pulsed mode) could be used with about $<$20\% of pulses resulting in signal on a given LAPPD channel. Of note, each pixel will have differences in gain as the quantum efficiency will differ across the LAPPD.
The Single-Photoelectron (SPE) data from an oscilloscope is shown in Fig.~\ref{fig_SPE_gain_TEKscope_laser_splitter_H1}; peak gain of about $6 \times 10^{6}$ was measured:
\begin{equation}
    G = \frac{Q_{SPE}}{e \; Z} 
    \simeq \frac{50\times 10^{-12} \;\mathrm{mV\; ns} }{1.6 \times 10^{-19} \;\mathrm{C} \ 50\; \Omega} 
    = 6 \times 10^{6}
\end{equation}

\begin{figure}
    \centering
    \includegraphics[width=1\linewidth]{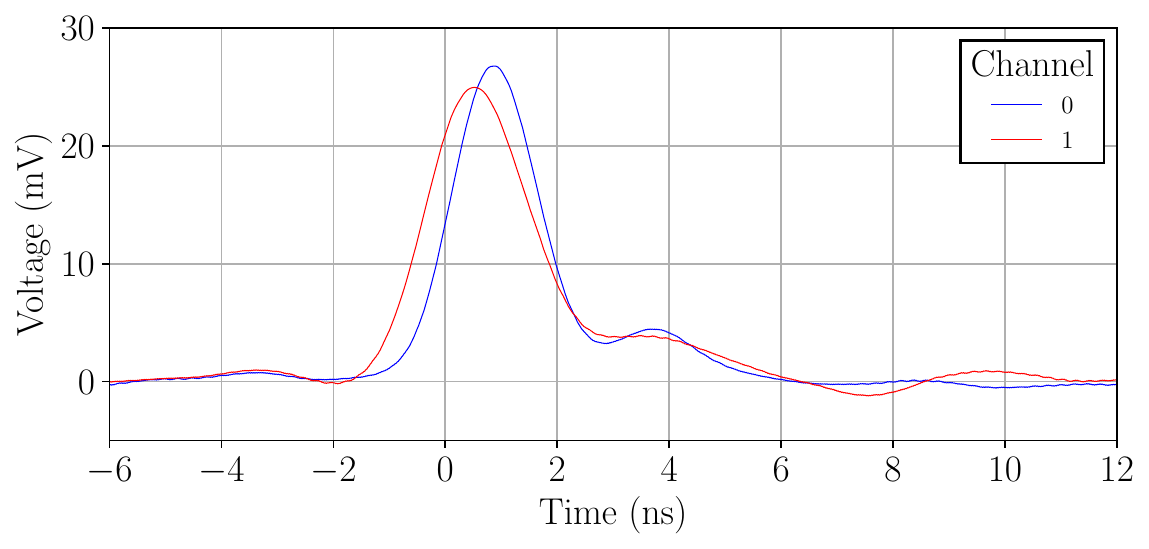}
    \caption{LAPPD waveforms from pixels H1 (channel 0) and E5 (channel 1) corresponding to a single-photoelectron level. Triggering is on delay generator pulse which also sends trigger to the laser controller. This is a raw data from an oscilloscope capture (Tektronix MSO64B).}
    \label{fig_2ch_SPE_TEKscope_laser_splitter_H1_E5}
\end{figure}

\begin{figure}
    \centering
    \includegraphics[width=1\linewidth]{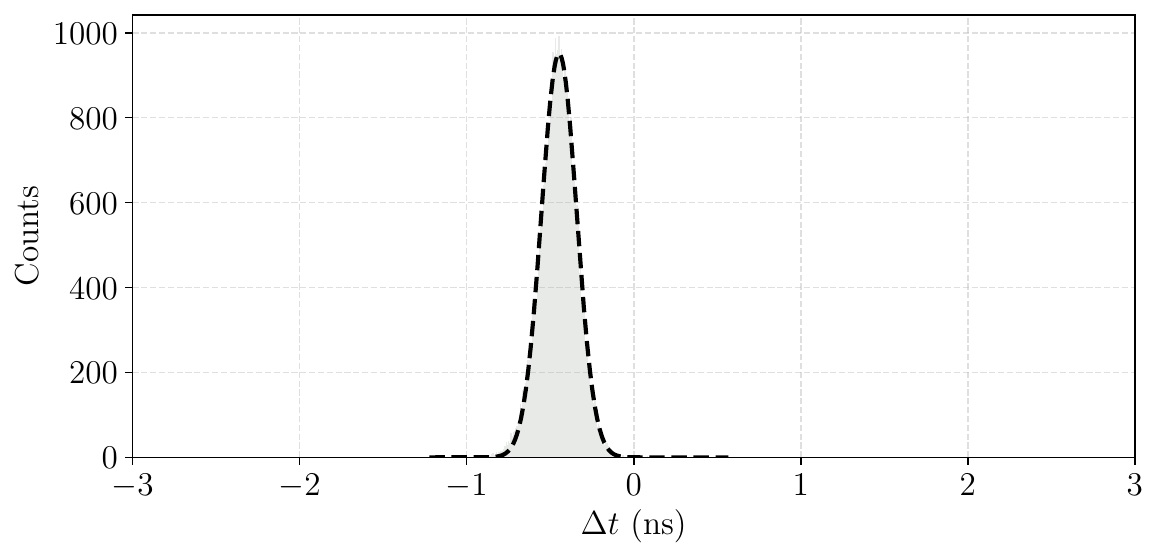}
    \caption{The time delay measurement with beam splitter laser at multi-PE levels on both pixels E5 and H1; this measurement is with Tektronix oscilloscope (at 25 GSa/s). The Gaussian mean is -446~ps, with a sigma of 105~ps. }
    \label{fig_jitter_TEKscope_laser_splitter_E5_H1}
\end{figure}

The laser used in this setup is Picoquant DL-800D with a 450-nm laser head (LDH-P-C).
Experimental measurements with a Thorlabs DET025AFC fast photodiode shows an inherent free-space laser pulse jitter of 5-10~ps. This pulse width and jitter are further broadened as the light propagates through the fiber and subsequently passes through a 50:50 single-mode fused-fiber splitter. Insufficient coupling efficiency limited our ability to measure the fiber output jitter; however, it is expected to have small contributions on the order of 10~ps. Utilizing this beam splitter, it is possible to estimate the timing jitter between a pair of LAPPD pixels. This jitter represents the convolution of contributions from optical jitter, electronics jitter, and the intrinsic LAPPD jitter. We measure dark count rate on the order of 100~Hz per pixel, using the oscilloscope at a trigger level of 4 mV.  
The measurement for a pair of pixels, H1 and E5, is presented in Figs.~\ref{fig_2ch_SPE_TEKscope_laser_splitter_H1_E5} and~\ref{fig_jitter_TEKscope_laser_splitter_E5_H1}. The observed delay of 446~ps corresponds to the electronic path trace separation between the two pixels subtracted by the physical separation between the movable collimator and the fixed collimator relative to the LAPPD face, as well as difference in trace length in the LAPPD PCB. 

\section{Readout of LAPPD Gen 2 with AARDVARC}

Following the successful initial assessment of the LAPPD Gen 2 using a fast-timing oscilloscope, we proceeded to integrate it with novel SoC readouts. For this study, we utilized two evaluation boards from Nalu Scientific: one based on the HDSoC and the other on the AARDVARC. The key features of these two chips are summarized in Table~\ref{tab_ASIC_spec}. Briefly, the HDSoC offers high channel density (32 channels in the version used), while the AARDVARC provides superior timing resolution. Although we operated the LAPPD with both systems, this report focuses on the AARDVARC due to its high sampling rate and suitability for fast-timing applications.

\begin{figure}
    \centering
     \begin{overpic}[width=1\linewidth]{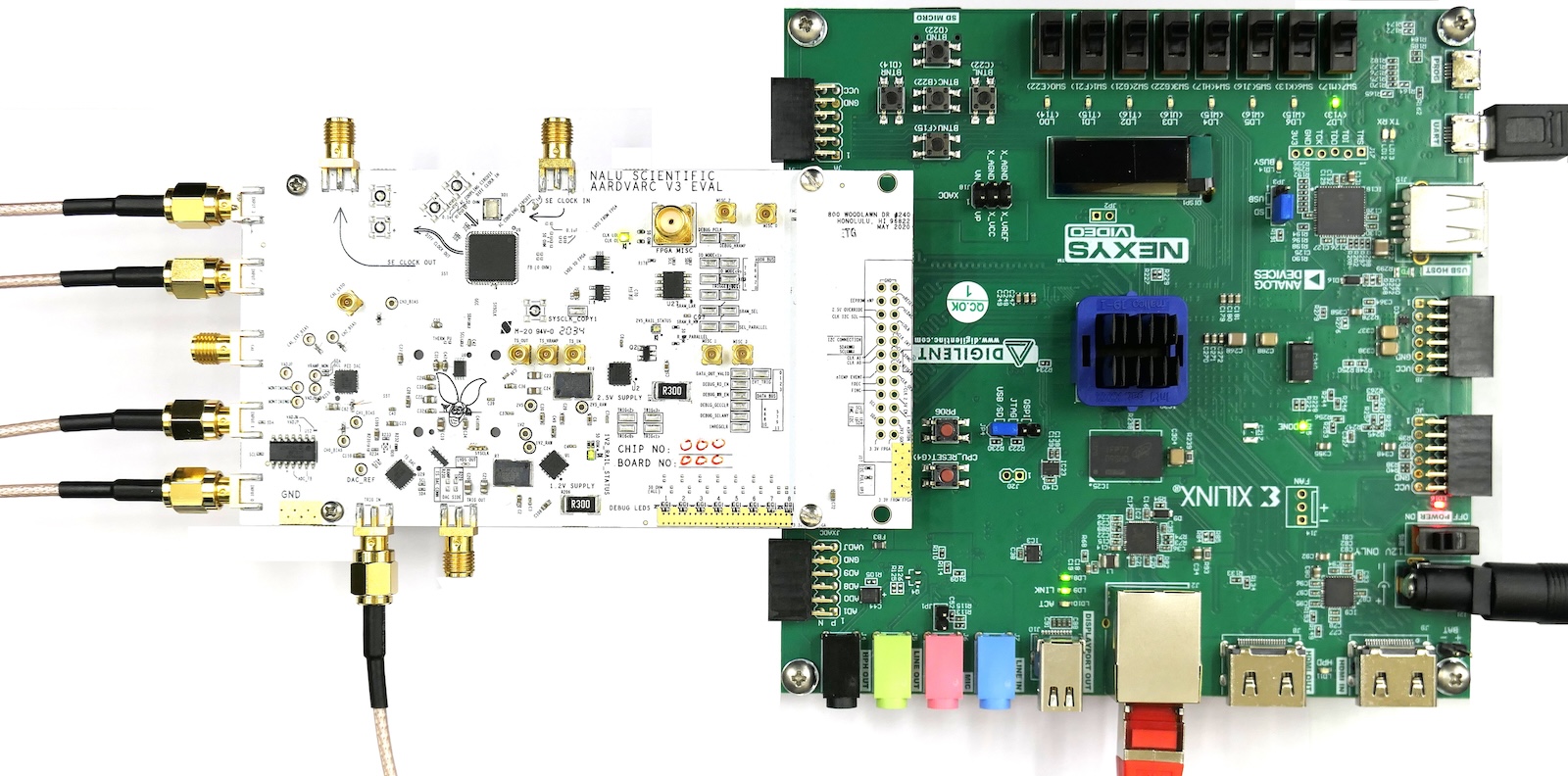} 
        \thicklines
        \put(12,5){\color{red}\vector(1,1){10}}
        \put(0,5){\fcolorbox{red}{white!30}{\parbox{30pt}{External trigger input}}}
        
        \put(13,47){\color{red}\vector(0,-1){8}}
        \put(0,45){\fcolorbox{red}{white!30}{\parbox{60pt}{4 input signals}}}

        \put(60,1){\color{red}\vector(1,0){11}}
        \put(45,0){\fcolorbox{red}{white!30}{\parbox{35pt}{Ethernet}}}

        \put(96,0){\color{red}\vector(0,1){10}}
        \put(85,0){\fcolorbox{red}{white!30}{\parbox{25pt}{Power}}}

        \put(97,33){\color{red}\vector(0,1){6}}
        \put(93,31){\fcolorbox{red}{white!30}{\parbox{10pt}{\tiny USB}}}
        
        \thinlines
    \end{overpic}
    \caption{The AARDVARC eval board (white) connected to the Nexys board (green) based on the AMD-Artix-7 FPGA.  
    The external trigger input and four ``signal'' channels have SMA cables connected in this photograph. The data collection and communication is done via Ethernet port on the Nexys card.}
    \label{fig_AARDVARC_eval}
\end{figure}

The Nalu evaluation board featuring the 4-channel AARDVARC system-on-a-chip is connected to a Nexys board, equipped with AMD Artix-7 FPGA, as illustrated in Fig.~\ref{fig_AARDVARC_eval}. The board communicates with a computer via Ethernet, and the ``NaluScope'' software (Windows version) is used to configure data acquisition parameters, trigger settings, and pedestal collection. 
The system supports both internal triggers (e.g., signals from any channel) and external triggers (e.g., signals from a delay generator).
Notably, the evaluation board used in this study bears the serial number ``000,'' underscoring the preliminary nature of this investigation.

Since our primary interest lies in evaluating how this photosensor and readout system would perform in a neutrino detector filled with water or water-based scintillator—where multiple photons are unlikely to hit the same small area—we focus on the timing response to multiple hits within the sensor, rather than within a single pixel. We tested the board in both ``external'' and ``signal'' trigger mode.

\begin{figure}[ht!]
    \centering
    \includegraphics[width=1\linewidth]{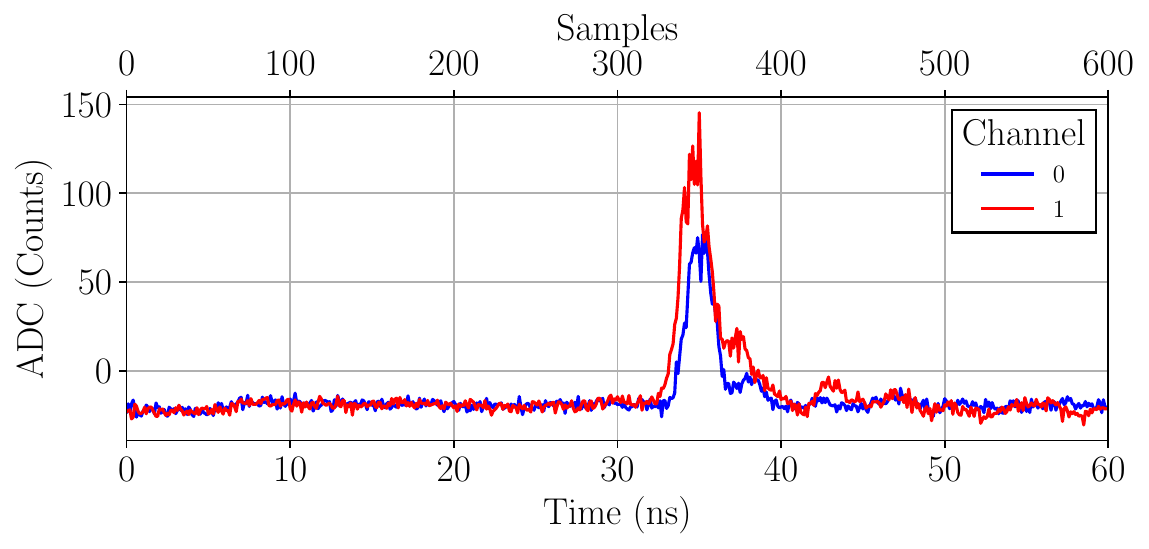}
    \caption{PE waveforms as obtained with AARDVARC system-on-a-chip connected to a pair of LAPPD channels (the output of anode H1 is connected to channel 0; E5 --- channel 1). }
    \label{fig_AARDVARC_H1_E5_LAPPD}
\end{figure}

\begin{figure}[ht!]
    \centering
    \includegraphics[width=1\linewidth]{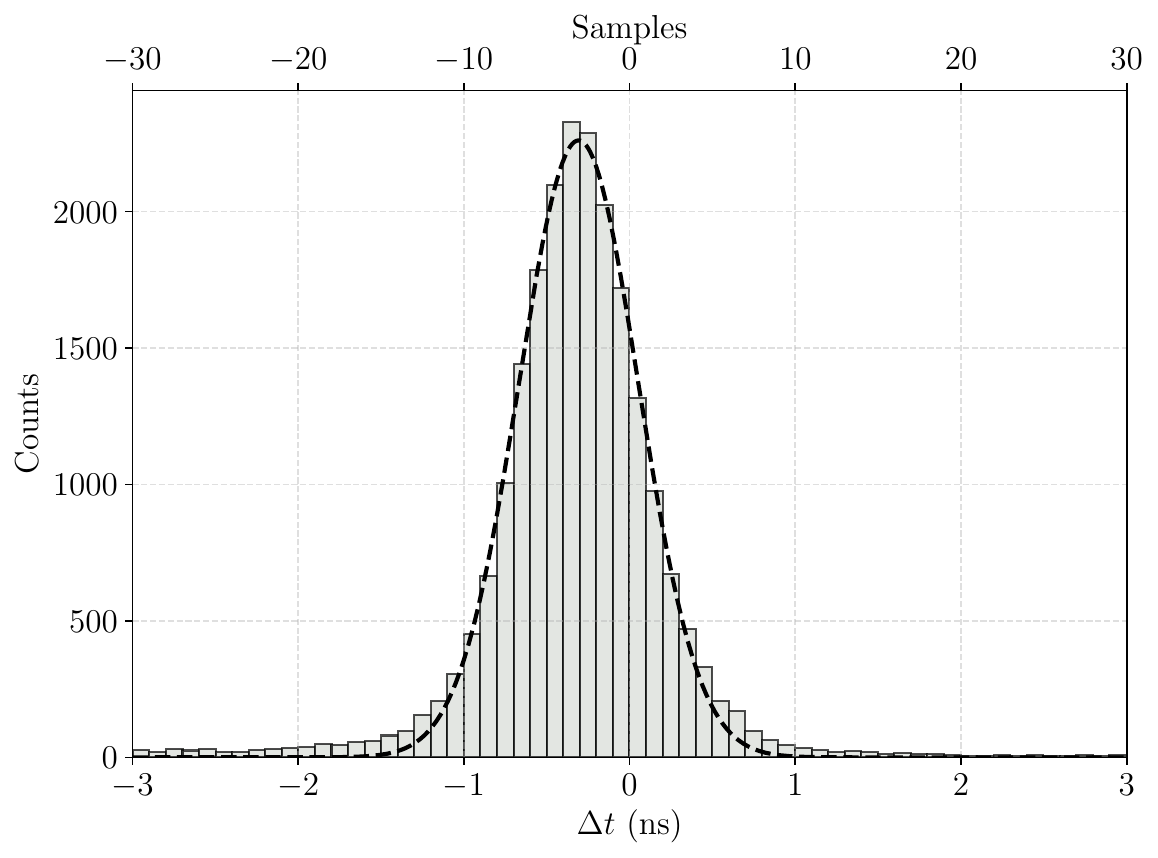}
    \caption{The delay measurement of the beam splitter laser at multi-PE levels on both pixels E5 and H1; this measurement is with AARDVARC on the internal trigger setting. The Gaussian mean and sigma are -308~ps and 360 ps, or -3.08 and 3.60 samples respectively.}
    \label{fig_jitter_AARDVARC_laser_splitter_E5_H1}
\end{figure}

To measure the timing delay between a pair of LAPPD pixels, as in the oscilloscope tests, we directed a laser through the beam splitter at two specific pixels (E5 and H1). The laser intensity was adjusted to operate in the SPE regime. An example of two SPE pulses from the corresponding LAPPD pixels is shown in Fig.~\ref{fig_AARDVARC_H1_E5_LAPPD}.
A log-normal fit was applied to the waveforms to estimate the rising edge position (defined as the mean minus sigma). The NaluScope software on AARDVARC does not contain a logic trigger setting for "AND" readout on both channels; therefore, waveform data was rejected from timing calculations if the fitted amplitude on both channels did not exceed a threshold value of 30 ADCs, the software unit equivalent to electric potential. This ensures that timing delay calculations are performed only on measurements with pulses detected on both channels and filters out false positives from background noise. In measurements, up to 80$\%$ of the raw waveform data was triggered on only one single channel and had to be rejected, which shows how the difference in quantum efficiency between the two pixels impacts AARDVARC data collection. The distribution of the time differences between the two pixels is presented in Fig.~\ref{fig_jitter_AARDVARC_laser_splitter_E5_H1}. Note the conversion rate of 1~sample to 100~ps. 

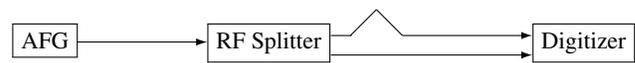
\begin{figure}[ht]
\setlength{\unitlength}{.2\linewidth}

\begin{picture}(5,0.3)

\put(0.5,0.05){\vector(1,0){1}}
\put(2.45,-0.05){\vector(1,0){1.55}}
\put(2.45,0.1){\line(1,0){0.16}}
\put(3,0.1){\line(-1,1){0.2}}
\put(2.6,0.1){\line(1,1){0.2}}
\put(3,0.1){\vector(1,0){1}}

\put(0,0){\fbox{AFG}}
\put(1.5,0){\fbox{RF Splitter}}

\put(4,0){\fbox{Digitizer}}

\end{picture}

\caption{The schematic of the setup to study electronic jitter.}
\label{fig_splitter_dia}
\end{figure}

\begin{figure}[ht!]
    \centering
    \includegraphics[width=1\linewidth]{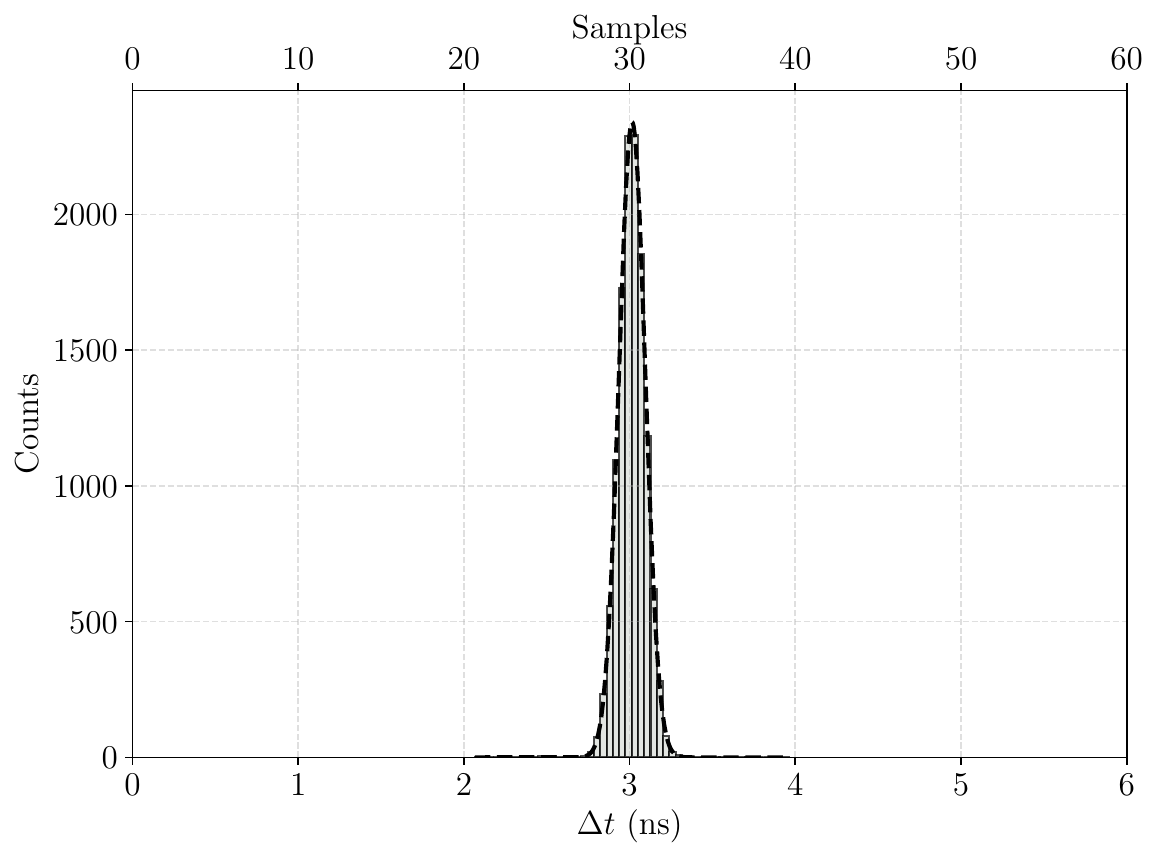}
    \caption{Delay measurement of AARDVARC with electrical SPE-like pulses at a known delay of 3 ns, using the internal signal trigger at n $\approx$ 10,000 events. The fitted Gaussian mean is 3.015 ns, with a sigma of 79 ps. When the pulses are triggered on the external trigger connected to the delay generator, the fitted Gaussian changes to a mean of 2.901 ns and a sigma of 51 ps.}
    \label{fig_AARDVARC_time_delay}
\end{figure}

The total measured timing resolution is given by the quadrature sum of the individual contributions:
\begin{equation}
    \sigma^2 =  \sigma_{\mathrm{laser}}^2 + \sigma_{\mathrm{LAPPD}}^2 + \sigma_{\mathrm{readout}}^2 
\end{equation}
To estimate the electronics contribution, we injected a pulse from a function generator (Tektronix AFG31252) into a two-way power splitter (Minicircuits ZAPD-30-S+), producing a signal shape similar to the LAPPD SPE pulse. The signal was modeled with a double-exponential function (40-mV high, and about 2-ns wide). The resulting waveforms were recorded using an oscilloscope, yielding a measured timing jitter of $<30$~ps (Gaussian sigma of the distribution), as shown in Fig.~\ref{fig_splitter_dia}. The observed delay corresponds to the known cable length of 3~ns. The channel-to-channel delay was measured using the Tektronix built-in delay measurement functionality.
A comparable measurement was performed using the Nalu AARDVARC readout, with results shown in Fig.~\ref{fig_AARDVARC_time_delay}. As with the laser/LAPPD measurements, a log-normal fit was applied to the waveforms. The standard deviation $\sigma$ of the timing distribution was measured to be $79$~ps.

\begin{table}[]
    \centering
    \begin{tabular}{|l|r|r|}\hline
       Specification & HDSoC  & AARDVARC \\\hline
       Channels & 32/64 & 4/8 \\
       Timing resolution & $<$100 ps & $<5$ ps (at 13 GSa/s) \\
       Sampling rate & 1-3 GSa/s & 10--14 GSa/s  \\
       Analog bandwidth & 1\;GHz & $>1.6$\;GHz \\
       Buffer length (Sa/ch.) & 2048 & 32k \\
       Trigger buffer & $\sim$2\;$\mu$s & $\sim$3\;$\mu$s \\
       Max rate zero-deadtime & 23 kHz/channel \footnote{full
readout, 200kHz hits/array} & 125 kHz  \\

       Supply voltage / range & 2.5\;V / 0.5--2.2\;V & 1.2\;V / 0.3--0.9\;V \\
       Input noise & 1 mV &  \\
       ADC bits & 12 & 12 \\
       Technology & 250 nm CMOS & 130 nm CMOS \\
       Power/channel & 20--40\;mW & 80\;mW \\\hline
    \end{tabular}
    \caption{Main specification for HDSoC and AARDVARC, from Nalu datasheets~\cite{AARDVARC2020,AARDVARC2023,HDSOC2023}, reproduced here for clarity.}
    \label{tab_ASIC_spec}
\end{table}

\section{Single-Photon Counting with LAPPD Gen 2}

\begin{figure}[ht!]
    \centering
    \includegraphics[width=1\linewidth]{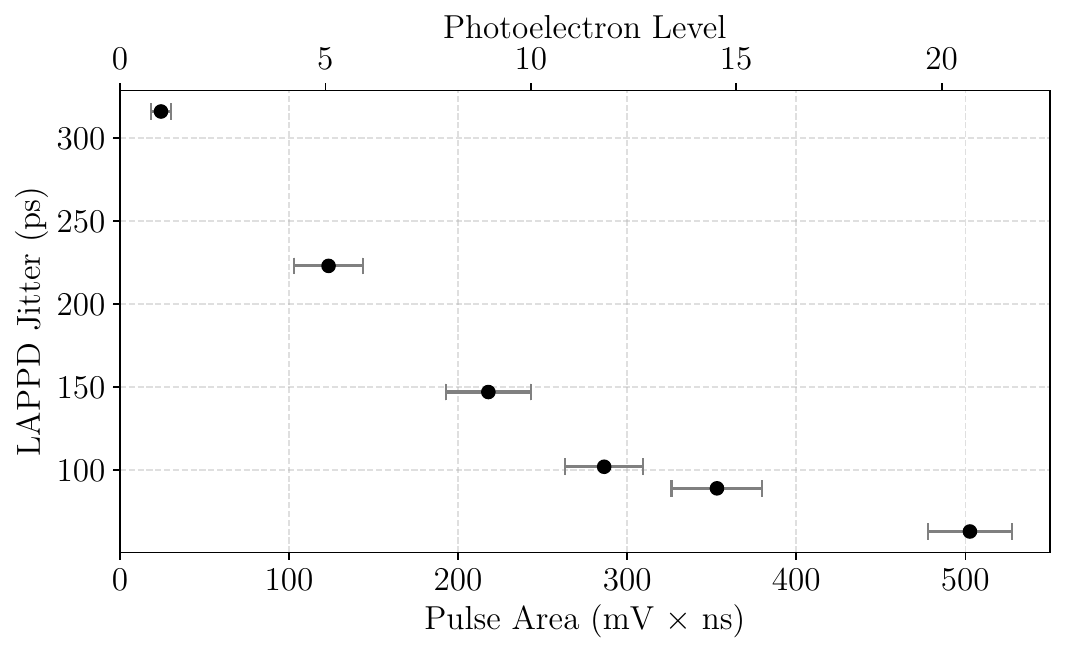}
    \caption{Measurement of the LAPPD jitter ($\sigma$) between E5 and G3 with respect to a change in the photoelectron level (adjusted by selecting neutral-density filters, keeping the laser intensity constant). Pixel E5 was measured for the total charge in this figure and has a measured SPE charge of 24.3 mVns.}
    \label{fig_jitter_vs_charge}
\end{figure}

At the conclusion of initial measurements, we noticed a rather large timing spread on the LAPPD that varied in ranges above the sub-100~ps jitter that was observed, which led to further investigations. 
With LAPPD measurements, we discovered that the timing delay and jitter between two pixels has a dependence on the incoming photon intensity. These variables were observed to worsen when laser intensity decreases towards SPE levels, as well as when one approaches the saturation limit for LAPPD. Peak temporal resolution occurs with high photoelectron levels as seen in Fig.~\ref{fig_jitter_vs_charge}. With this in mind, we redesigned the experiment to test the inherent time delay and jitter between two pixels while minimizing external signal delay influences and maintaining an intensity that corresponds to SPE levels. 

\begin{figure}[ht]
\setlength{\unitlength}{.2\linewidth}

\begin{picture}(5,0.6)
\put(0.72,0.55){\color{blue}\vector(1,0){0.75}}
\put(2.75,0.55){\color{blue}\vector(1,0){1.25}}
\put(2,0.4){\color{blue}\vector(0,-1){0.2}}
\put(2.45,0.05){\color{blue}\vector(1,0){1.55}}
\put(0,0.5){\fbox{LASER}}
\put(1.5,0.5){\fbox{Optical Splitter}}
\put(1.8,0){\fbox{Mirror}}
\put(4,0){\rotatebox{90}{\fbox{LAPPD}}}
\end{picture}

\caption{The schematic of the setup to study LAPPD jitter without fibers.}
\label{fig_optical_splitter_LAPPD_mirror}
\end{figure}
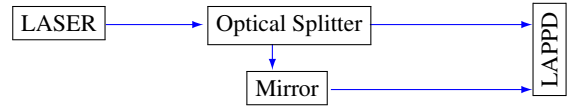

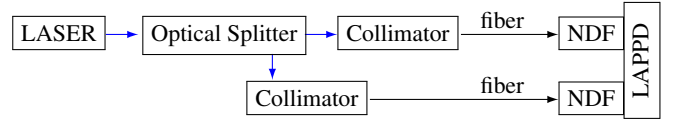
\begin{figure}[ht]
\setlength{\unitlength}{.2\linewidth}

\begin{picture}(5,1)
\put(0.72,0.55){\color{blue}\vector(1,0){0.25}}
\put(2.25,0.55){\color{blue}\vector(1,0){0.25}}
\put(2,0.4){\color{blue}\vector(0,-1){0.2}}
\put(2.75,0.05){\color{black}\vector(1,0){1.45}}
\put(3.45,0.55){\color{black}\vector(1,0){0.75}}
\put(0,0.5){\fbox{LASER}}
\put(1.,0.5){\fbox{Optical Splitter}}
\put(2.5,0.5){\fbox{Collimator}}
\put(1.8,0){\fbox{Collimator}}
\put(4.2,0){\fbox{NDF}}
\put(4.2,.5){\fbox{NDF}}
\put(4.7,-.1){\rotatebox{90}{\fbox{{\color{white}\_}LAPPD{\color{white}\_}}}}
\put(3.6,.1){fiber}
\put(3.6,.6){fiber}
\end{picture}

\caption{The schematic of the setup to study LAPPD jitter  using fibers and combinations of neutral density filters (labelled as "NDF" on the diagram). Note that the two fibers are roughly the same length and not shown to scale ($\sim$2~m length for both fibers).}
\label{fig_optical_splitter_LAPPD_fibers_and_NDFs}
\end{figure}

\begin{figure}[ht!]
    \centering
    \includegraphics[width=1\linewidth]
    {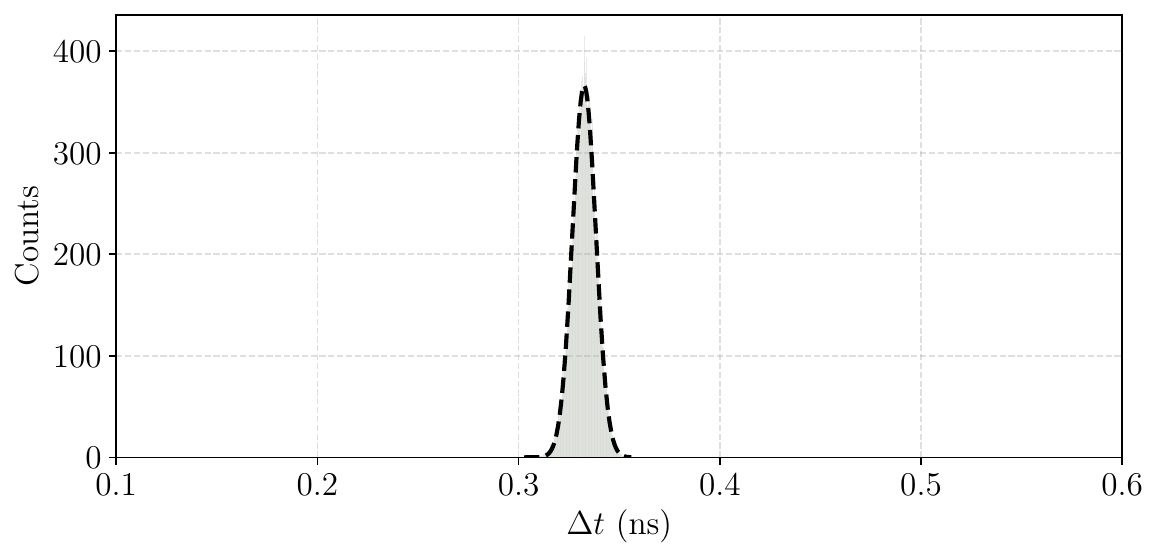}
    \caption{Measurement of the timing delay and jitter in the two-fiber setup made with two photodiodes at high laser intensity. The Gaussian mean is at 333 ps, with a sigma of 5.8 ps.}
    \label{fig_fiberjitter}
\end{figure}

Two experimental setups were constructed to measure the time delay between pixels in the LAPPD. The first setup consists of a free-space optical laser setup where a 50:50 beamsplitter directs one beam onto one pixel while the second "dump" beam uses a mirror to target a second pixel. Figure~\ref{fig_optical_splitter_LAPPD_mirror} shows this setup, and was also used to couple into two photodiodes to measure the inherent laser jitter between two pulses as noted in previous sections. The second setup combines this optical laser setup with the earlier fiber setup, where the laser and beamsplitter couples into two Thorlabs FG050LGA fibers, which are then directed onto two pixels, as seen in Fig.~\ref{fig_optical_splitter_LAPPD_fibers_and_NDFs}. Measurement of laser jitter at the output of the two fibers is shown in Fig.~\ref{fig_fiberjitter}.

In both setups, the initial laser beam intensity is kept at a high level to maintain the low jitter that was measured with the photodiodes, so neutral density filters were used to attenuate the beam / fiber output to SPE levels at the output. We confirm that we achieve SPE levels after lowering the signal amplitude to 15-20 mV by comparing the trigger rate of the LAPPD, set by the logic trigger "AND" on two channels that correspond to the targeted pixels on the oscilloscope, with the delay generator trigger rate and background levels, where we measured $\sim$300-500~Hz compared to the pulse 10~kHz trigger rate. 
Integrating the pulse area for one SPE measured $\sim$24.3 mV\;ns on pixel E5 to further confirm SPE levels, which is seen with the curve on Fig.~\ref{fig_jitter_vs_charge}.

\begin{figure}[ht!]
    \centering
    \includegraphics[width=1\linewidth]{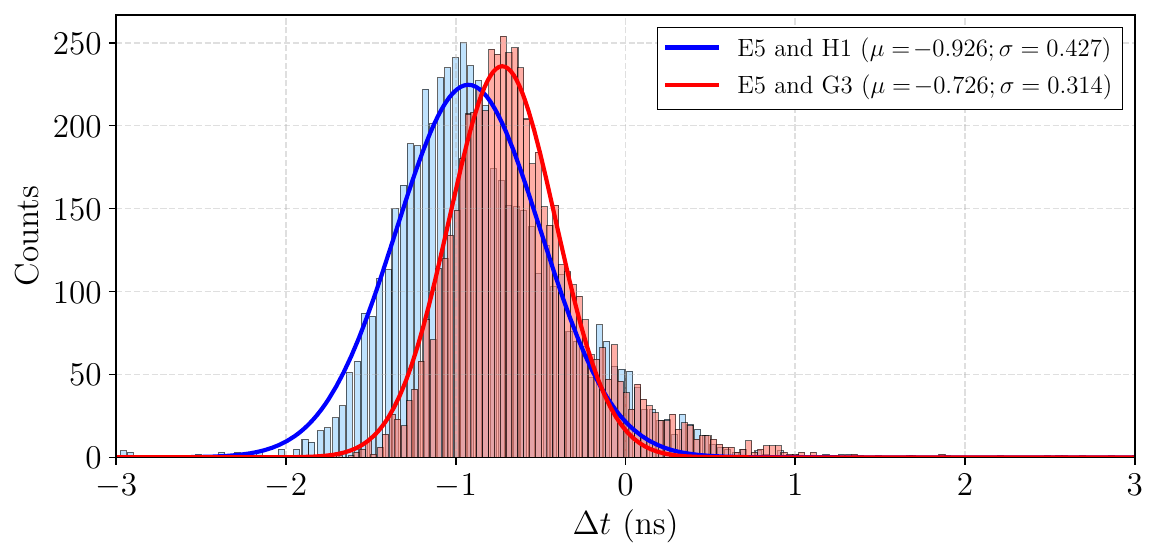}
    \caption{Measurement of the timing delay and jitter in the LAPPD with two fibers on the oscilloscope between two pairs of pixels at SPE levels. This shows how the jitter and timing delay remains large, but also slightly shortens based on pixel path distance. Similar results were measured by the free-space laser setup.}
    \label{fig_LAPPDjitter_fiber}
\end{figure}

\begin{figure}[ht!]
    \centering
    \includegraphics[width=1\linewidth]{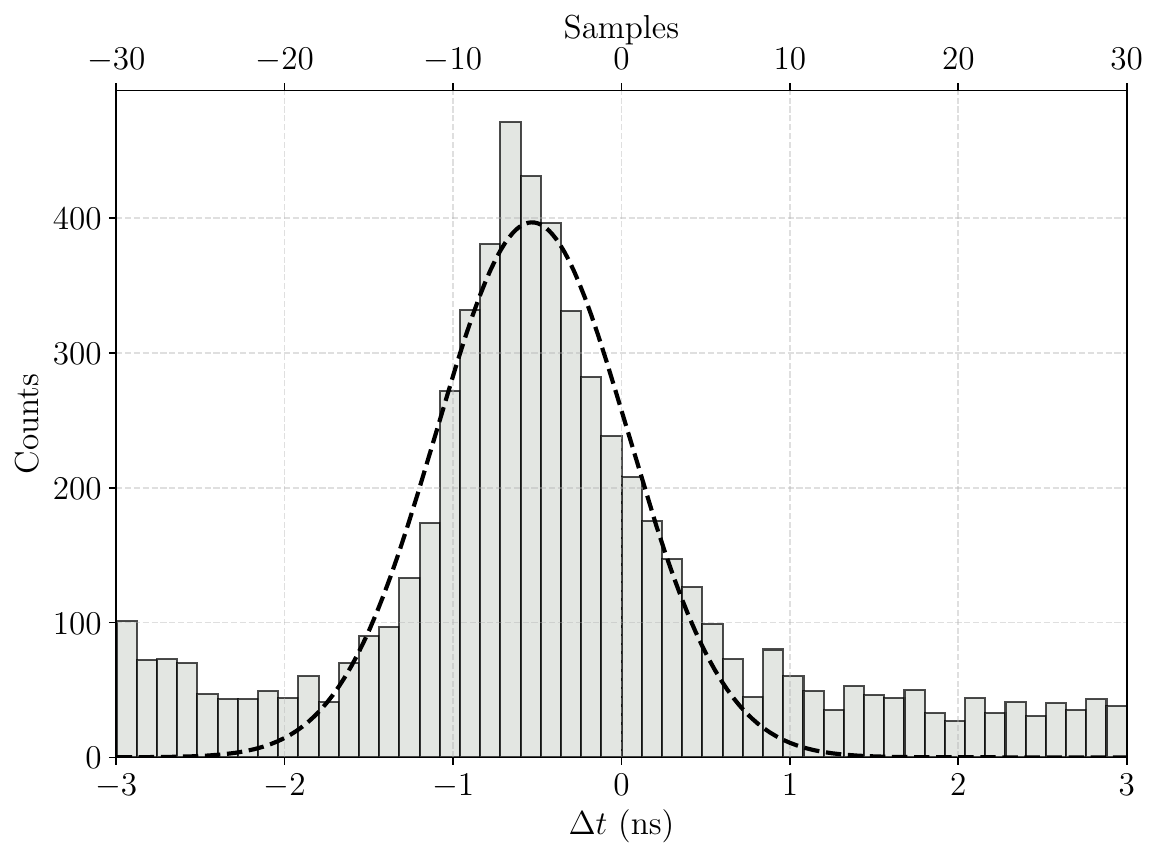}
    \caption{AARDVARC measurement of E5 and G3 pixel delay. The fitted Gaussian mean is -531~ps, with a sigma of 570~ps, reproducing results that agree with measurements from the Oscilloscope in Fig~\ref{fig_LAPPDjitter_fiber}. Notably for AARDVARC, there is a high background of $\sim$50-100 counts outside of the expected distribution. This issue is likely due to a combination of fitting issues within distinguishing SPE pulses against electronic background noise, and not using the full calibration capabilities of AARDVARC.}
    \label{fig_LAPPDjitter_AARDVARC}
\end{figure}

Multiple measurements performed with a free-space laser and the two-fiber setup shows that the LAPPD has a 2-pixel signal jitter of up to 500~ps, depending on the pixel pair chosen. Two measured pairs are shown in Figure~\ref{fig_LAPPDjitter_fiber}. Similar measurements were also performed with AARDVARC, as seen in Figure~\ref{fig_LAPPDjitter_AARDVARC} with a 20\% detection rate at the trigger for both pixels which obtains timing data that agrees with the oscilloscope. 
At the SPE level, noise becomes appreciable in AARDVARC two-pixel delay measurements, yielding counts outside the expected timing distribution. Full NaluScope configuration with stricter filtering and improved fit criteria should mitigate, and potentially eliminate, SPE-level erroneous counts.

The signal delay observed by both AARDVARC and the oscilloscope aligns well with the expected value due to the PCB trace difference and the laser air path difference. 
Our measurements indicate that, for the tested configuration with two-pixel coincidence timing at SPE level, the jitter is generally above 100 ps. We note however, that in some applications it might not be critical. Sub-100 ps timing was achieved for individual pixels relative to an external reference or at multi-photoelectron levels, suggesting good performance in externally triggered or higher-signal regimes, as seen in Fig.~\ref{fig_jitter_vs_charge}.

\section{Impact and Future Work}

This report presents the first successful integration of the LAPPD Gen 2 with the AARDVARC and HDSoC systems-on-a-chip. Initial testing has demonstrated promising performance for both the new photosensors and the electronics readout system. This versatile setup enables detailed studies of cross-talk and timing jitter, offering valuable insights for further integration. Importantly, the potential applications of this readout system extend beyond large neutrino detectors, making it relevant to a wide range of scientific endeavors. We hope this work will be of interest to the broader scientific community, as it continues a series of LAPPD-related and compact fast-timing readout studies published in Review of Scientific Instruments~\cite{LAPPD2013,LAPPD2019,LAPPD2020, mTC:2016yys}.

Since the introduction of the second generation of the LAPPD~\cite{Shin:2022ybc}, several groups have reported studies on various anode geometries and readout configurations for diverse applications~\cite{Seljak:2022khz, Bhattacharya:2023nmj, Xie:2024vzl, Barnyakov:2024zei, Agarwala:2024mfr, Dolenec:2024qkg}. In low-light-yield environments, such as pure water or WbLS, the timing performance of the detector becomes critical, particularly when single photons striking the photocathode in close proximity produce signals across multiple pads on the same photosensor. Multi-anode MCP-PMTs exhibit intrinsic cross-talk effects that require further investigation to determine such impacts on pixel-by-pixel signal output~\cite{mTC_MCP_2019}. On the readout side, improving pedestal temperature corrections is likely to reduce electronic noise.

The setup presented in this report provides a platform for conducting comprehensive signal response studies, which will be used to characterize cross-talk effects by sending a laser signal targeting a single pixel and measuring the cross-talk response in neighboring pixels. The amplitude and total charge of the pulses in neighbor pixels will be evaluated with respect to the real signal pulse in the target pixel, along with their responses relative to the position of the incoming laser. Near-term studies that are in-progress at the time of writing involve raster scanning across the entire $8\times8$ pixel pad array to develop a comprehensive timing study that characterizes path-trace timing differences, as well as cross-talk responses in neighboring pixels from spatial dependence of the input laser on the target pixel. Deployment of the LAPPD with radiation sources,  scintillators, and Cherenkov detectors could be done in order to characterize the signal response from physical events. These experiments will be combined with simulations on the LAPPD MCP electron avalanche to determine the radial distribution of photoelectrons after contact with the anode along with another simulation on coupled capacitance between pixels. This has the goal of producing a theory that describes the signal produced by the LAPPD that matches experimental data.

\section{Acknowledgments}

This work was performed under the auspices of the U.S. Department of Energy by Lawrence Livermore National Laboratory, Livermore, CA, USA under Contract DE-AC52-07NA27344. LLNL-JRNL-872766.

This research was supported in part by an appointment to the National Nuclear Security Administration Minority Serving Institutions Internship Program (NNSA-MSIIP), sponsored by the U.S. Department of Energy and administered by the Oak Ridge Institute for Science and Education.

We thank technical support of Tektronix, Nalu Scientific, and Incom for fruitful discussions and assistance.
The work of J.F. and S.-W. S. was partially supported by DOE's Reaching New Energy Workforce Initiative.

\section{Data Availability}
The data that support the findings of this study are available
from the corresponding author upon reasonable request.

\bibliography{ref}
\bibliographystyle{aipnum4-2}

\onecolumngrid

\end{document}